\documentclass[
superscriptaddress,
amsmath,
amssymb,
12pt,
reprint,
floatfix]
{revtex4-2}
\usepackage{amsthm}
\usepackage{bibunits}
\defaultbibliographystyle{unsrt}
\usepackage[utf8]{inputenc}
\usepackage{newtxtext}
\usepackage{newtxmath}
\usepackage[colorlinks=true,linkcolor=magenta]{hyperref}
\usepackage{physics}
\usepackage{here}
\usepackage{mathtools}
\usepackage{bm}
\usepackage[english]{babel}
\usepackage{color}
\usepackage{comment}
\usepackage[utf8]{inputenc}
\usepackage{color}
\usepackage[pgf]{adjustbox}

\usepackage{ulem}
\usepackage{fancybox}
\usepackage{amsmath,amssymb,amsfonts,mathrsfs}

\usepackage{graphicx}
\newcommand{\mc}[1]{\mathcal{#1}}
\newcommand{\mr}[1]{\mathrm{#1}}

\newcommand{\mb}[1]{\mathbb{#1}}
\newcommand{\msf}[1]{\mathsf{#1}}

\newcommand{\wtil}[1]{\widetilde{#1}}
\newcommand{\blue}[1]{\textcolor{blue}{#1}}
\newcommand{\red}[1]{\textcolor{red}{#1}}

\theoremstyle{remark}

\begin{document}
\title{Fractionalization paves the way to local projector embeddings of quantum many-body scars
}
\author{Keita Omiya}
\affiliation{Department of Physics, ETH Z\"{u}rich, CH-8093 Z\"{u}rich, Switzerland}
	\affiliation{Condensed Matter Theory Group, LTC, SCD, Paul Scherrer Institute, Villigen PSI CH-5232, Switzerland}
	\affiliation{Institute of Physics, Ecole Polytechnique F\'ed\'erale de Lausanne (EPFL), CH-1015 Lausanne, Switzerland}
	\author{Markus M\"uller}
\affiliation{Condensed Matter Theory Group, LTC, SCD, Paul Scherrer Institute, Villigen PSI CH-5232, Switzerland}
	\begin{abstract}
		Many systems that host exact quantum many-body scars (towers of energy-equidistant low entanglement eigenstates) are governed by a Hamiltonian that splits into a Zeeman term and a sum of local terms that annihilate the scar subspace. We show that this unifying structure also applies to models, such as the Affleck-Kennedy-Lieb-Tasaki (AKLT) model or the PXP model of Rydberg-blockaded atoms,
        that were previously believed to evade this characterisation. To fit these models within the local annihilator framework we need to fractionalize  their degrees of freedom and enlarge the associated Hilbert space. The embedding of the original system in a larger space elucidates the structure of their scar states and simplifies their construction, revealing close analogies with lattice gauge theories. 
	\end{abstract}
\maketitle
\section{Introduction}\label{sec:intro}
Interacting systems hosting a large set of exceptional eigenstates with equidistant energies, so-called  quantum many-body scars (QMBS), have attracted a lot of interest as they constitute a new class of ergodicity-breaking systems. In generic quantum many-body systems, eigenstates with the same energy density are believed to be thermodynamically equivalent in the sense that the quantum mechanical expectation values of local observables are independent of the eigenstate and equal to those obtained in the micro-canonical ensemble. This eigenstate thermalization~\cite{Deutsch1991,Srednicki1994,Deutsch2018_review,Rigol2008} arises due to the maximally possible entanglement caused by interactions in states of finite energy density above the ground state. 
Systems violating the eigenstate thermalization hypothesis (ETH) are rather rare, but are of particular interest, since unlike thermalizing system they are able to store information and may preserve a memory of their initial conditions. This includes integrable systems, where an extensive number of conserved quantities inhibits equilibration to the standard Gibbs ensemble. In many-body localized  (MBL) systems~\cite{Gornyi2005,Basko2006,Abanin2019}, instead strong disorder results in quasi-local conserved quantities, so-called local integrals of motion (LIOMs)~\cite{PhysRevB.90.174202, ROS2015420, Imbrie2016}, that hinder equilibration. While in these examples the dynamics is non-ergodic for arbitrary initial states, systems hosting 
QMBS exhibit a more subtle ergodicity-breaking: it is only seen in the tiny fraction of the Hilbert space that manages to evade ETH. However, once the system is initialized within the subspace of the exceptional scar states, an exactly time-periodic motion results.

Originally an experiment on a chain of Rydberg atoms  suggested such a possibility~\cite{Bernien2017}. Although this system is neither integrable nor in an MBL phase, unusually long-lasting oscillations were observed when the system was initialized in certain specific density wave states. It is by now well-established that these oscillations arise due to a set of special and weakly entangled states having almost equidistant energies~\cite{Turner2018Nature,Turner2018}. 
Since then intensive research on QMBS has uncovered a variety of non-integrable models that host scar states, including the Hubbard model~\cite{eta_pairing, eta_pair_rsga, PhysRevB.102.075132, https://doi.org/10.48550/arxiv.2205.07235} and the Affleck-Kennedy-Lieb-Tasaki (AKLT) model~\cite{Moudgalya2018, aklt_nonthermal}, and many more are discussed in the review~\cite{Sanjay2021Review}. 

In addition to identifying sets of scar states in specific models~\cite{Moudgalya2018,Schecter2019,Iadecola2020DW,PhysRevLett.123.136401, PhysRevB.101.245107, PhysRevB.102.180302, PhysRevX.11.021008, PhysRevResearch.3.L012010, PhysRevB.102.195150, PhysRevLett.126.220601}, several recipes to generate further models with such properties have been proposed~\cite{ShiraishiMori2017, Mark2020,topological_scar,Kiryl2020,PhysRevB.102.085120, ODea2020,Kiryl2021, Ren2021}. 
The known recipes broadly fall into two classes: One is based on identifying a set of ``annihilator'' terms in the Hamiltonian that annihilate the entire scar subspace~\cite{ShiraishiMori2017,Kiryl2020,PhysRevB.102.085120, ODea2020, Ren2021}. The other class is based on the so-called spectrum generating algebra (SGA)~\cite{eta_pairing, Mark2020, PhysRevX.12.011050}. Both methods aim at fragmenting the Hilbert space into dynamically disconnected subspaces. In the annihilator approach, local operators fragment the Hilbert space into their kernel and its complement. QMBS Hamiltonians can then be obtained by taking a simple  integrable Hamiltonian $H_{\mr{spec}}$ with an equidistant spectrum and adding to it a sum of annihilating terms, $H_{\mr{ann}}$.
On the complement of the common kernel of the annihilators appearing in $H_{\mr{ann}}$ the spectrum is generically chaotic, such that $H$ is non-integrable. However, those eigenstates of $H_{\mr{spec}}$ that lie in the subspace of annihilated states remain eigenstates of $H$ and retain an equidistant spectrum. $H_{\mr{ann}}$ can usually be expressed as a sum of local terms, each of which annihilates the subspace of scar states. We refer to such operators as local annihilators. They form an  algebra~\cite{https://doi.org/10.48550/arxiv.2209.03377}, and the scar subspace can be characterized as the intersection of their kernels. An instructive example of this construction is the projector embedding (or the Shiraishi-Mori form)~\cite{ShiraishiMori2017}, where the Hamiltonian is a sum of terms $P_i h_i P_i$, where the $P_i$ are local projectors that annihilate the scar states and the $h_i$ are arbitrary Hermitian operators. 
Note that the notion of {\em locality} of the annihilators is important. Indeed, if the scar subspace can be characterized as the intersection of their kernels, this suggests that it might be possible to represent it by a superposition of a small number of suitably constrained tensor network states with a low bond dimension.

In contrast to the above construction, the SGA constructs disconnected subspaces by non-local ``creation'' operators $Q^+$ (that are usually extensive, translationally invariant sums of local terms). These creation operators have to be such that, when restricted to a certain subspace $\mc{W}$,
they satisfy the equation of motion of a harmonic oscillator, i.e., $[H,Q^+]\mc{W}=\omega Q^+\mc{W}$. A tower of scar states with equidistant spacing $\omega$ can then be obtained by applying the creation operators multiple times to a reference eigenstate state in $\mc{W}$. The resulting states resemble excited states with multiple, infinitely long-lived  quasi-particle excitations.

It remains a challenge to reconcile the local annihilator approach and the SGA. The AKLT model of a spin 1 chain, for example, is known to host exact scar states~\cite{aklt_nonthermal} that can be understood via the SGA. Furthermore, Mark \textit{et.al.}, found that the AKLT model can indeed be decomposed into a spectrum generating Zeeman term and an annihilating term~\cite{Mark2020}. However, they also discovered that the latter cannot be expressed as a sum of local annihilators, which led them to conclude that the AKLT model should be considered ``beyond'' the local annihilator formalism, implying potentially that these scars might not admit a compact characterization in terms of tensor networks.
Moudgalya and Motrunich recently generalized the annihilator approach, using the notion of the bond algebra~\cite{PhysRevX.12.011050,https://doi.org/10.48550/arxiv.2209.03377} generated by the different local terms in the Hamiltonian. Considering its irreducible representations and the center of the algebra (the ``commutant'') as well as using the bi-commutant theorem, they managed to characterize the scar subspace, and confirmed that only non-local operators can take the role of annihilators $H_{\mr{ann}}$ in the AKLT model. The non-local annihilators can be chosen to be extensive sums of local terms that are invariant under certain translations. However, since the local terms are not by themselves annihilators, these operator sums are   sometimes called ``as-a-whole'' annihilators.

{The absence of local annihilators in this and certain other models raises the question of whether there are truly different classes of QMBS. Moreover, the non-locality of annihilators is an unsatisfactory property, as it appears to hamper an efficient description of scar states in terms of tensor network states.}

A system where several of the above issues arise, too, is the original Rydberg chain. It is described by the so-called PXP model, a deep understanding of which still presents a challenge. 
Finite-size diagonalization shows that the PXP model hosts non-thermal scar states, but they only form an approximate tower whose energy spectrum is not exactly equidistant~\cite{Turner2018Nature}. So far no exact form for these states is known, and with the exception of a few selected states no simple closed expression is expected to capture them, in contrast to many models with exact scar states. Therefore, the thermodynamic limit of the PXP is not entirely settled, although there are strong indications that the scar states do violate the ETH in the thermodynamic limit~\cite{PhysRevB.98.155134}. 

The scar states of the PXP model have been explored in numerous papers, primarily using the SGA approach~\cite{Choi2019, Iadecola2019pimagnon}. However, their wavefunctions could not be obtained within this approach, hindering analytical predictions of physical observables. In a recent paper~\cite{PhysRevA.107.023318} we filled this gap by showing that these scar states can largely be understood within a (generalized) local annihilator approach. A key insight is that, due to its constrained Hilbert space (the conservation of the number of simultaneously excited neighboring atoms), the annihilators are not local in the original Hilbert space. However, by considering an enlarged Hilbert space and defining a proper embedding of the model, one can obtain a scarred Hamiltonian with local annihilators.

In this paper, we will show that this structure generalizes to several spin models, for which hitherto a local projector embedding was not known.
 In particular, we show that non-local annihilators, such as those of the AKLT model, can indeed be decomposed as a sum of \textit{local} terms, provided the model is embedded into an appropriately enlarged Hilbert space. 
Since there is no unique way to lift a Hamiltonian to an enlarged Hilbert space, one can take advantage of the remaining freedom in such a way that the new Hamiltonian has the desired decomposition into $H_{\mr{spec}}$ and a sum of local annihilators $H_{\mr{ann}}$. Our approach rationalizes in a unifying manner the emergence of scar states in both, the conventional projector-embedding, and in models with ``as-a-whole'' annihilators. Moreover, it establishes a relation to the mechanism at play in the PXP model. {Our construction of scars has a lot of resemblance with structures appearing in lattice gauge theories (LGTs). Those possess a ``physical configuration space'' defined by Gauss' law, which is embedded in an enlarged space containing non-physical states. Ground states of LGTs often take a very simple form in the enlarged space, once a suitable gauge is fixed. This is closely analogous to the simpler form the scar states take in a suitably enlarged Hilbert space.
}

The remainder of the paper is organized as follows. We briefly review  
the projector embedding as an example of the local annihilator approach and introduce our notation in Sec.~\ref{sec:toy model}. 
In Sec.~\ref{sec:PXP} we set the stage by revisiting the construction of exact scar states for a suitably modified PXP Hamiltonian, comparing various ways of extending the Rydberg-constrained Hilbert space.
In Sec.~\ref{sec:spin_xy}, we first consider the spin-1 XY model which hosts two different towers of scar states.
We show that one of them is characterized by local annihilators, while the other one, the tower of ``bi-magnon'' states, only possesses  hidden local annihilators, requiring an enlargement of the Hilbert space to obtain local annihilators. In Sec.~\ref{sec:AKLT}, we show that the AKLT model, too, has a very similar structure. We finally discuss the spin-$1/2$ domain-wall conserving model in Sec.~\ref{sec:dw model}. There we show that by considering the full Hilbert space as an extension of its subspaces with fixed domain wall number leads to natural understanding of the
 various towers of ``pyramidal'' states and furnishes an interesting analogy with the topological degeneracy appearing in LGTs. In Sec.~\ref{sec:lgt}, we discuss the similarity between the projective construction of  ground states in the deconfined phase of LGTs and their analogue for scar states discussed in Secs.~\ref{sec:spin_xy}$\sim$\ref{sec:dw model}.

\section{Local annihilator approach}\label{sec:toy model}
\subsection{The projector embedding}
We start by reviewing the projector embedding as a simple example of the local annihilator approach. It splits the Hamiltonian into two parts $H=H_{\mr{spec}}+H_{\mr{ann}}$, whereby $H_{\mr{ann}}$ is assumed to be a sum of local annihilators of the form 
\begin{equation}\label{eq:projector embedding}\begin{split}
    &H_{\mr{ann}}=\sum_i\mc{P}_ih_i\mc{P}_i\\
    &\left[H_{\mr{spec}}, \mc{P}_i\right]=0,
\end{split}
\end{equation}
where $\mc{P}_i$ is a local projector and $h_i$ is a local Hermitian operator, both labelled by lattice site $i$. The scar states $\ket*{\mc{S}_n}$ belong to the subspace $\mc{W}=\bigcup_i\ker\mc{P}_i$, defined as the  intersection of the kernels of $\mc{P}_i$, i.e., $\mc{P}_i\ket*{\mc{S}_n}=0$. We note that the existence of a  non-trivial intersection is usually not obvious, as the projectors do not commute in general. The second condition in Eq.~\eqref{eq:projector embedding} ensures that $H_{\mr{spec}}$ leaves $\mc{W}$ invariant, as $\mc{P}_iH_{\mr{spec}}\mc{W}=H_{\mr{spec}}\mc{P}_i\mc{W}=0$. This entails that the eigenstates of $H_{\mr{spec}}$ that lie in $\mc{W}$ form a set  of energy equidistant states.

The form of $H_{\mr{ann}}$ can be generalized if we do not require the Hamiltonian to be Hermitian. In that case, the most general form of $H_{\mr{ann}}$ is $H_{\mr{ann}}=\sum_ih_i\mc{P}_i$ where $h_i$ is an arbitrary local operator. We will see that such non-Hermitian annihilators often appear when the Hilbert space is enlarged.

\subsection{Hilbert space extension and lifting of QMBS Hamiltonians}\label{sec:main idea}
Here we will demonstrate that the scar states in the PXP model, the bi-magnon states in the spin-1 XY model and the scar states in the AKLT model can all be understood in a similar way, namely as a generalized local projector embedding, even though their original scar spaces do not admit such a structure.  
For a given Hamiltonian $H$, we construct an embedding of the Hilbert space $\cal H$ into a larger space $\wtil{\cal H}$, i.e., a map $\msf{P}$ which projects $\wtil{\cal H}$ onto ${\cal H}$. In the enlarged space, we then construct a lifted Hamiltonian $\wtil{H}$ satisfying 
\begin{equation}\label{eq:desired decomp}
    H\msf{P}=\msf{P}\wtil{H},
\end{equation}
 such that it can be decomposed as a spectral part and a sum of local annihilators,
\begin{equation}\label{eq:desired decomp}\begin{split}
    H\msf{P}&=\msf{P}\wtil{H},\\
    \wtil{H}&=\wtil{H}_{\mr{spec}}+\wtil{H}_{\mr{ann}}.
    \end{split}
\end{equation}
The lifted Hamiltonian $\wtil{H}$ now admits scar states $\ket*{\wtil{\mc{S}}_n}$, within the standard local annihilator paradigm.
The corresponding scar states of the original model are finally obtained upon projection as $\ket*{{\mc{S}}_n}= \msf{P}\ket*{\wtil{\mc{S}}_n}$.

As {we will show in the subsequent sections}, it is important that the choice of $\wtil{H}$ is not unique. This redundancy defines an equivalence class of operators $\wtil{A}$ and $\wtil{B}$ in the enlarged space. We consider them equivalent, writing $\wtil{A}\sim\wtil{B}$, if $\msf{P}\wtil{A}=\msf{P}\wtil{B}$ holds. Clearly, $\wtil{A}\sim\wtil{B}$ and $\wtil{C}\sim\wtil{D}$ imply $\wtil{A}+\wtil{C}\sim\wtil{B}+\wtil{D}$. While often there is a ``natural'' lifted Hamiltonian $\wtil{H}_{\mr{nat}}$ in the enlarged space, it typically does not have the desired decomposition.
 We thus need to seek a suitable equivalent operator $\wtil{H} \sim \wtil{H}_{\mr{nat}}$ having a decomposition~\eqref{eq:desired decomp}. 

From an alternative viewpoint one may regard $\msf{P}$ as a map from operators in the enlarged space to those in the original space. We should thus seek a suitable preimage of this map satisfying Eq.~\eqref{eq:desired decomp}, which can be formally written as $\wtil{H}\in\msf{P}^{-1}H$.

In the following, we will show the explicit construction of $\msf{P}$ and $\wtil{H}$ for {the scar states in the PXP model}, the bi-magnon scars in the spin-1 XY model and the scar states in the AKLT model. Most importantly, the resulting annihilators for those models are local in the extended Hilbert space. In other words, those models can be cast into a generalized projector embedding from, i.e., 
\begin{equation}\label{eq:generalized pe}
    H\msf{P}=\msf{P}\left(\wtil{H}_{\mr{spec}}+\sum_i\wtil{h}_i\wtil{P}_i\right),
\end{equation}
where $\wtil{h}_i$ is some local operator and $\wtil{P}_i$ is a local projector in the enlarged Hilbert space. Scar states $\ket*{\wtil{\mc{S}}_n}$ are annihilated by $\wtil{P}_i$ for $\forall i$. 
{In all these cases $\wtil{H}_{\mr{spec}}$ is a simple Zeeman term that commutes with the $\wtil{P}_i$, thus fulfilling the requirements of a projector embedding (Eq.~\eqref{eq:projector embedding}).} Our construction via a Hilbert space extension provides a way to make the annihilators local, even in models that were believed to only admit long-ranged annihilators for their scar subspaces.

\section{The PXP model}\label{sec:PXP}
We follow~\cite{PhysRevA.107.023318} to review the PXP model as an explicit example for a generalized local projector embedding, based on a map that embeds  the Rydberg constrained Hilbert space into a space of larger dimension. 

The PXP model is a one-dimensional spin-1/2 model defined as
\begin{equation}
    H_{\mr{PXP}}=\sum_{i\in\Lambda}P_{i-1}\sigma^x_iP_{i+1},
\end{equation}
where $P\coloneqq\dyad*{\downarrow}$ is the projector onto the down spin and the $\sigma^\alpha (\alpha=x,y,z)$ are the Pauli matrices. We assume an even number of sites  $N\coloneqq|\Lambda|$ and periodic boundary conditions, $i+N\equiv i$.

\subsection{Emergent spin-1 formulation}
This model commutes with the Rydberg projector $P_{\mr{Ryd}}\coloneqq\prod_{i\in\Lambda}(1-\dyad*{\uparrow\uparrow})_{i,i+1}$, which prohibits simultaneous excitation of neighboring spins. Within the subspace ${\cal R}$ satisfying this constraint, there are only three possible configurations of two neighboring spins, ($\ket*{+}\equiv\ket*{\downarrow\uparrow}$, $\ket*{-}\equiv\ket*{\uparrow\downarrow}$, and $\ket*{0}\equiv\ket*{\downarrow\downarrow}$). Upon choosing a dimerization of the chain, one can identify these states with a basis of a pseudo-spin 1. In this basis  $H_{\mr{PXP}}$ takes the  form 
\begin{equation}\label{eq:PXP s-1}\begin{split}
    \wtil{H}_{\mr{PXP}}&\coloneqq\sqrt{2}\sum_{b\in\Lambda_{\mr{B}}}S_b^x+\wtil{H}',\\
    \wtil{H}'&\coloneqq-\sum_{b\in\Lambda_{\mr{B}}}\ket*{+,-}\left(\bra*{+,0}+\bra*{0,-}\right)_{b,b+1}+h.c.\\
    &\equiv -\wtil{O}'-\wtil{O}'^\dagger,\\ \wtil{O}'&\coloneqq\sum_b\ket*{+,-}(\bra*{+,0}+\bra*{0,-})_{b,b+1},
     \end{split}
\end{equation}
where $\Lambda_{\mr{B}}=\{1,\cdots,N_b\equiv N/2\}$ is the lattice of dimers, or (pseudo) block-spins, $S^x$ being the  associated spin-1 operator. The Rydberg projector in the spin-1 space can be written as $P_{\mr{Ryd}}\coloneqq\prod_{b\in\Lambda_{\mr{B}}}(1-\dyad*{+,-})_{b,b+1}$. The full spin-1 space thus takes the role of an enlarged Hilbert space that contains the Rydberg manifold ${\cal R}$ as a submanifold.

By our construction, the above defined spin-1 model and the original PXP Hamiltonian (restricted to ${\cal R}$) satisfy the following relation, 
\begin{equation}
\label{eq:lifting}
   \begin{split} &H_{\mr{PXP}}\msf{P}_{\mr{Ryd}}=\msf{P}_{\mr{Ryd}}\wtil{H}_{\mr{PXP}}\\
   &\msf{P}_{\mr{Ryd}}\coloneqq\left[\prod_{b\in\Lambda_{\mr{B}}}\left(\dyad*{\downarrow\uparrow}{+}+\dyad*{\uparrow\downarrow}{-}+\dyad*{\downarrow\downarrow}{0}\right)_b\right]P_{\mr{Ryd}},
   \end{split}
\end{equation}
where $\msf{P}_{\mr{Ryd}}$ is the Gutzwiller-type projection from the full spin-1 space to ${\cal R}$. 
As for the notation, $\dyad*{\downarrow\uparrow}{+}_b$ should be read as $\equiv\ket*{\downarrow\uparrow}_{2b-1,2b}\bra*{+}_b$, the first index pair ($2b-1,2b$) referring to the original chain sites, in contrast to the dimer index $b$.

We emphasize that $\wtil{H}'$ in Eq.~\eqref{eq:PXP s-1} is \textit{not} a sum of local annihilators, not even approximately. However, 
since $\msf{P}_{\mr{Ryd}} \wtil{O}'=0$, the modified lifted Hamiltonian $\wtil{H}_{\mr{PXP}}\to \wtil{H}_{\mr{PXP}}+\wtil{O}'$ still satisfies Eq.~\eqref{eq:lifting}. 

The piece which is added to the Zeeman term in this modified Hamiltonian,  namely $\wtil{H}'+\wtil{O}' \equiv  -\wtil{O}'^\dag$, turns out to approximately consist of local annihilators. To turn this into an exact statement, one may further add to the original Hamiltonian the following non-Hermitian perturbation (cf.~\cite{PhysRevA.107.023318})
\begin{equation}\label{eq:non-H perturbation}
    \delta H_{\mr{NH}}\coloneqq\frac{1}{2}\sum_{b\in\Lambda_{\mr{B}}}\left(\ket*{+,0}+\ket*{0,-}\right)\bra*{0,0}_{b,b+1}
\end{equation}
which commutes with the Rydberg projector. Upon lifting the perturbed Hamiltonian $H+\delta H_{\mr{NH}}$ to the enlarged Hilbert space, it can be decomposed into the Zeeman term (providing an equidistant spectrum) and a sum over exact local annihilators,
\begin{equation}\label{eq:s1 mod PXP}\begin{split}
    &\wtil{H}_{\mr{PXP}}+\wtil{O}'+\delta H_{\mr{NH}}=\sqrt{2}\sum_{b\in\Lambda_{\mr{B}}}S_b^x+\sum_{b\in\Lambda_{\mr{B}}}h_b\left(1-P^{S=2}_{b,b+1}\right).
    \end{split}
\end{equation}
Here, $P^{S=2}_{b,b+1}$ is the two-body projector onto the sector of total spin $2$. The scar subspace, which is annihilated by all those projectors, is the space of maximal total spin, and the exact scar eigenstates of this perturbed PXP model take the form $\ket*{\mc{S}^{\mr{PXP}}_n}\equiv\msf{P}_{\mr{Ryd}}\ket*{\wtil{S}_n}$ where $\ket*{\wtil{S}_n}$ is an eigenstate of the total $S^x$ operator, within the sector of maximal spin. 

This construction is appealing as it makes explicit the picture that the scars of the PXP model arise from a precessing large spin. Note, however, that due to the choice of a dimerization both the lifted Hamiltonian $\wtil{H}_{\mr{PXP}}$ and the parent state $\ket*{S_n}$ break translational symmetry (by a single spin-1/2 site). However, by projecting back onto the Rydberg manifold, the physical scar states $\ket*{\mc{S}^{\mr{PXP}}_n}(=\msf{P}_{\mr{Ryd}}\ket*{S_n})$ turn out to be independent of the choice of dimerization and  invariant under translation, as we will explain next.

\subsection{Translation invariant formulation}
To understand the resulting translation invariance of the scar state $\ket*{\mc{S}^{\mr{PXP}}_n}$, we now present an equivalent construction, which is however based on a translation invariant lift of the Hamiltonian, that again takes the form of Eq.~\eqref{eq:desired decomp}. Instead, it will not have an equally natural interpretation in terms of a large spin.

We consider the non-Hermitian perturbed PXP model
\begin{equation}
    \label{eq:non-Hermitian PXP}
    H_{\mr{nH-PXP}}\coloneqq H_{\mr{PXP}}+\frac{1}{4}\sum_{i\in\Lambda}P_{i-1}\sigma_i^+P_{i+1}\left(P_{i-2}+P_{i+2}\right),
\end{equation}
where $\sigma^+\coloneqq\dyad{\uparrow}{\downarrow}$ is the raising operator, and the second term is representing $\delta H_{\mr{NH}}$ (Eq.~\eqref{eq:non-H perturbation}) averaged over translations. (Due to the translation invariance of the scar states, they are exact eigenstates of the translation symmetrized Hamiltonian, too).

$H_{\mr{nH-PXP}}$ also commutes with the Rydberg projector, and satisfies  similar equations as Eqs.~(\ref{eq:PXP s-1},\ref{eq:lifting}),
\begin{equation}
    \label{eq:trsl inv PXP in unconstrained}
    \begin{split}
        &H_{\mr{nH-PXP}}P_{\mr{Ryd}}=P_{\mr{Ryd}}\left(\sum_{i\in\Lambda}\sigma_i^x+H'_{\mr{nH}}\right),\\
        &H'_{\mr{nH}}=-\frac{1}{2}\sum_{i\in\Lambda}\left(n_{i-1}\sigma_i^x+\sigma_i^xn_{i+1}+P_{i-1}\sigma_i^xn_{i+1}+n_{i-1}\sigma_i^xP_{i+1}\right)\\
        &\qquad+\frac{1}{4}\sum_{i\in\Lambda}P_{i-1}\sigma_i^+P_{i+1}\left(P_{i-2}+P_{i+2}\right),
    \end{split}
\end{equation}
where $n\coloneqq\dyad*{\uparrow}= 1-P$ is the projector onto the up-spin. We now look for an operator equivalent to $H'_{\mr{nH}}$, which consists of local annihilators. Using the identities $n_{i-1}\sigma_i^+\sim_{\mr{Ryd}}\sigma_i^+n_{i+1}\sim_{\mr{Ryd}}0$ and $\sigma_i^+P_{i+1}\sim_{\mr{Ryd}}P_{i-1}\sigma_i^+\sim_{\mr{Ryd}}\sigma_i^+$ (with the equivalence relation $A\sim_{\mr{Ryd}}B\Leftrightarrow P_{\mr{Ryd}}A=P_{\mr{Ryd}}B$),
we obtain
\begin{equation}
 \begin{split}
    H'_{\mr{nH}}&\sim_{\mr{Ryd}}  -\frac{1}{2}\sum_{i\in\Lambda}\left(n_{i-1}\sigma_i^-+\sigma_i^-n_{i+1}+P_{i-1}\sigma_i^- n_{i+1}+n_{i-1}\sigma_i^-P_{i+1}\right)\\
        &\qquad+\frac{1}{4}\sum_{i\in\Lambda}\sigma_i^+P_{i+1}P_{i+2} +P_{i-2}P_{i-1}\sigma_i^+,
   \end{split}
\end{equation}
which, upon a reorganization of the three-site terms, yields the equivalent operator
\begin{equation}\label{eq:modified H_nH}
    \begin{split}
        H'_{\mr{nH}}&\sim_{\mr{Ryd}}\wtil{H}'_{\mr{nH}}\\
        \wtil{H}'_{\mr{nH}}&=-\frac{1}{2}\sum_{i\in\Lambda}\left(\ket*{\uparrow\downarrow}+\ket*{\downarrow\uparrow}\right)\bra{\uparrow\uparrow}_{i,i+1}\\
        &-\frac{1}{2}\sum_{i\in\Lambda}\ket*{\downarrow\downarrow\uparrow}\left(\bra{\downarrow\uparrow\uparrow}-\frac{1}{2}\bra*{\downarrow\downarrow\downarrow}\right)_{i-1,i,i+1}\\
        &-\frac{1}{2}\sum_{i\in\Lambda}\ket*{\uparrow\downarrow\downarrow}\left(\bra{\uparrow\uparrow\downarrow}-\frac{1}{2}\bra*{\downarrow\downarrow\downarrow}\right)_{i-1,i,i+1}.
    \end{split}
\end{equation}

Note that the bra-part of the three-site terms contain factors $\bra{\uparrow\uparrow}-\frac{1}{2}\bra{\downarrow\downarrow}$. 
To bring the first line into a similar form, we would like to add and subtract a term $\frac{1}{4}\sum_{i}\left(\ket*{\uparrow\downarrow}+\ket*{\downarrow\uparrow}\right)\bra{\downarrow\downarrow}_{i,i+1}= \frac{1}{4}\sum_{i} (\sigma_i^+ P_{i+1} + P_{i-1}\sigma_i^+)\sim_{\mr{Ryd}} \sum_i \sigma_i^+/2$.
Adding a corresponding neutral term to the Zeeman term we find the equivalence
\begin{equation}\label{eq:modified Zeeman}
    \sum_{i\in\Lambda}\sigma_i^x\sim_{\mr{Ryd}}\sum_{i\in\Lambda}\left(\sigma_i^x-\frac{1}{2}\sigma_i^+\right)+\frac{1}{4}\left(\ket*{\uparrow\downarrow}+\ket*{\downarrow\uparrow}\right)\bra*{\downarrow\downarrow}_{i,i+1},
\end{equation}
Putting Eqs.~(\ref{eq:modified H_nH},\ref{eq:modified Zeeman}) together we finally find
\begin{equation}
    \begin{split}
        H_{\mr{nH-PXP}}&\sim_{\mr{Ryd}}H'_{\mr{spec}}+H'_{\mr{ann}}\\
        H'_{\mr{spec}}&=\sum_{i\in\Lambda}\sigma_i^x-\frac{1}{2}\sigma_i^+\equiv\sum_{i\in\Lambda}W_i\\
        H'_{\mr{ann}}&=-\frac{1}{2}\sum_{i\in\Lambda}\left(\ket*{\uparrow\downarrow}+\ket*{\downarrow\uparrow}\right)\left(\bra{\uparrow\uparrow}-\frac{1}{2}\bra{\downarrow\downarrow}\right)_{i,i+1}\\
        &-\frac{1}{2}\sum_{i\in\Lambda}\ket*{\downarrow\downarrow\uparrow}\left(\bra{\downarrow\uparrow\uparrow}-\frac{1}{2}\bra*{\downarrow\downarrow\downarrow}\right)_{i-1,i,i+1}\\
        &-\frac{1}{2}\sum_{i\in\Lambda}\ket*{\uparrow\downarrow\downarrow}\left(\bra{\uparrow\uparrow\downarrow}-\frac{1}{2}\bra*{\downarrow\downarrow\downarrow}\right)_{i-1,i,i+1}.
    \end{split}
\end{equation}

We are now ready to construct the scar states. 
Recall that we analyze a modified PXP Hamiltonian that was adjusted to guarantee that it hosts scar states. So our task is merely to identify the scar subspace by reverse engineering. It is natural to expect the scar subspace to be annihilated by all local terms $\bra{\uparrow\uparrow}-\frac{1}{2}\bra{\downarrow\downarrow}$, and to expect the remaining piece of the Hamiltonian,  $H'_{\mr{spec}}=\sum_i W_i$, to be the spectrum generating part.

To show that this is indeed the case we first diagonalize $H'_{\mr{spec}}$. The single-site operator $W$ have eigenvectors
\begin{equation}
    \begin{split}
        W\ket*{\pm}&=\pm\frac{1}{\sqrt{2}}\ket*{\pm}\\
        \ket*{\pm}&\coloneqq\pm\frac{1}{\sqrt{3}}\ket*{\uparrow}+\sqrt{\frac{2}{3}}\ket*{\downarrow}.
    \end{split}
\end{equation}
Note that $\ket*{\pm}$ are not orthogonal to each other since $W$ is non-Hermitian. We also define the associated dual vectors $\bra*{\pm'}$, requiring $\innerproduct*{\sigma'}{\tau}=\delta_{\sigma,\tau}$ with $\sigma,\tau=\pm$. 

We now claim that the scar states $\ket*{\wtil{\mc{S}}^{\mr{PXP}}_n}$ of the lifted Hamiltonian given above take the simple form
\begin{equation}\begin{split}
   \ket*{\wtil{\mc{S}}^{\mr{PXP}}_n}&=\frac{1}{\mc{N}'_n}\left(\sum_{i\in\Lambda}(-1)^i\dyad*{+}{-'}_i\right)^n\bigotimes_{j\in\Lambda}\ket*{-}_j,
    \end{split}
\end{equation}
which can be seen as $n$ excitations (of momentum $\pi$) on top of the fully site-symmetric reference state of all sites being in state $\ket*{-}$. An excitation consists in promoting one of the local $\ket*{-}$ states to $\ket*{+}$, acting on all sites in parallel, with an amplitude that alternates in space.

These states are easily seen to be eigenstates of $H'_{\mr{spec}}$ with eigenvalue $(-N/2+n) \sqrt{2}$.
Let us now show that in addition they are indeed all annihilated  by the local operators constituting $H'_{\mr{ann}}$.
This can be seen as follows. It follows from the alternating structure of the excitation operator that the possible local configurations of $\ket*{\wtil{\mc{S}}^{\mr{PXP}}_n}$ restricted to the sites $i, i+1$ are either $\ket*{\phi_-}\coloneqq\ket*{-,-},\ket*{\phi_0}\coloneqq(\ket*{+,-}-\ket*{-,+})/\sqrt{2}$, or $\ket*{\phi_+}\coloneqq\ket*{+,+}$. All these configurations are annihilated by the bra-part of $H'_{\mr{ann}}$. Indeed the latter was arranged so as to contain factors $\bra*{\uparrow\uparrow}-1/2\bra*{\downarrow\downarrow}\propto\bra*{+',-'}+\bra*{-',+'} \equiv \bra{X'}$, which is orthogonal to $\ket*{\phi_{+,0,-}}$ and thus annihilates $\ket*{\wtil{\mc{S}}^{\mr{PXP}}_n}$.


Note that the above procedure furnishes an explicit construction of the projector-embedding form of the modified PXP Hamiltonian in the unconstrained space,
\begin{equation}
    H_{\mr{nH-PXP}}P_{\mr{Ryd}}=P_{\mr{Ryd}}\left(\sum_{i\in\Lambda}W_i+\sum_{i\in\Lambda}h_i
    \dyad*{X}{X'}_{i,i+1}\right)
\end{equation}
where $h_i$ is some local operator. 
Note that now the scar states are manifestly invariant under a single-site translation, since the space enlargement and the lifting of the Hamiltonian preserved translation symmetry. This ensures that all its eigenstates and their Rydberg projections are translation invariant as well. 

\subsection{Relation between the two formulations}
We have shown that the scar state $\ket*{\mc{S}^{\mr{PXP}}_n}$ is obtained by seemingly different constructions based on a fictitious spin-1 lifted Hamiltonian and a (non-Hermitian) spin-1/2 Hamiltonian acting on the unconstrained Hilbert space of spin-1/2. Let us now discuss the relation between them.

Note first that the spin-1 space is naturally embedded in the unconstrained spin-1/2 space.  
One can then verify that up to annihilators the spin-1 Zeeman term  ($\sqrt{2}\sum_{b\in\Lambda_{\mr{B}}}S_b^x$) is equivalent to $H'_{\mr{spec}}(=\sum_{i\in\Lambda}W_i)$ in the spin-1/2 space. Indeed, a straightforward calculation yields
\begin{equation}\begin{split}
    &\sqrt{2}\sum_{b\in\Lambda_{\mr{B}}}S_b^x=\sum_{b\in\Lambda_{\mr{B}}}\left(\dyad*{\downarrow\uparrow}{\downarrow\downarrow}+\dyad*{\uparrow\downarrow}{\downarrow\downarrow}\right)_{2b-1.2b}+h.c.\\
    &\sim_{\mr{Ryd}}\sum_{i\in\Lambda}W_i-\sum_{b\in\Lambda_{\mr{B}}}\left(\ket*{\uparrow\downarrow}+\ket*{\downarrow\uparrow}\right)\left(\bra*{\uparrow\uparrow}-\frac{1}{2}\bra*{\downarrow\downarrow}\right)_{2b-1,2b},
    \end{split}
\end{equation}
which is a sum of $H'_{\mr{spec}}$ and local annihilators. The annihilator part in the spin-1 space ($-\wtil{O}'^\dag+\delta H_{\mr{NH}}$) can be rewritten as a sum of local annihilators in the spin-1/2 space, 
\begin{equation}
    \begin{split}
        &-\wtil{O}'^\dag+\delta H_{\mr{NH}}\\
        &=-\sum_{b\in\Lambda_{\mr{B}}}\left(\ket*{\downarrow\uparrow\downarrow\downarrow}+\ket*{\downarrow\downarrow\uparrow\downarrow}\right)\left(\bra*{\downarrow\uparrow\uparrow\downarrow}-\frac{1}{2}\bra*{\downarrow\downarrow\downarrow\downarrow}\right)_{2b-1;2b+2},
    \end{split}
\end{equation}
where the bra part again contains a local annihilator as a factor.

The parent scar states  $\ket*{S_n}$ in the spin-1 space, i.e. , the states of maximal total spin, are obtained from the parent states  $\ket*{\wtil{\mc{S}}^{\mr{PXP}}_n}$ in the unconstrained spin-1/2 space by imposing only half of the Rydberg constraints so as to enforce the spin-1 degrees of freedom on dimers,
\begin{equation}
    \ket*{S_n}=\frac{1}{\mc{N}'_n}\prod_{b\in\Lambda_{\mr{B}}}\left(1-\dyad*{\uparrow\uparrow}\right)_{2b-1,2b}\ket*{\wtil{\mc{S}}^{\mr{PXP}}_n}.
\end{equation}

We point out that the spin-1 lift of the PXP model is not always equivalent to a spin-1/2 lift. The most important difference is that the spin-1/2 lift can be applied to the PXP model on any bipartite lattice, while the spin-1 lift requires that the lattice has a dimer covering. This is not possible for many Rydberg systems including the real experimental setup of a 51-atom chain~\cite{Bernien2017}. For this reason, the spin-1/2 lift seems more fundamental to capture the hidden structure of the scar states of general PXP models, even though the spin-1 lift provides an explicit construction of a the ``large precessing spin precession'' that is alluded to in various papers (e.g., Ref.~\cite{Choi2019}).

\section{Spin-1 XY model}\label{sec:spin_xy}
An example of a system hosting scars with non-local annihilators in the original Hilbert space is the XY model for spins with $S=1$. We denote the $S=1$ basis that diagonalizes $S^z$
by $\ket*{\pm}$ and $\ket*{0}$, referring to the eigenvalues $\pm1$ and $0$, respectively. While this system hosts QMBS on hypercubic lattices in arbitrary dimensions~\cite{Schecter2019}, we restrict ourselves to the one-dimensional case for simplicity. The extension to higher dimensions is straightforward though. 
The Hamiltonian reads
\begin{equation}\label{eq:spin-1 XY}    H_{\mr{XY}}=J\sum_{i\in\Lambda}\left(S_i^xS_{i+1}^x+S_i^yS_{i+1}^y\right)+\sum_{i\in\Lambda}\left(hS_i^z+D\left(S_i^z\right)^2\right),
\end{equation}
where $\Lambda\coloneqq\{1,2,\cdots,N\}$ is a linear chain of sites, and the $S^\alpha (\alpha=x,y,z)$ are spin-1 operators. Here we assume periodic boundary conditions (PBC), identifying $i$ and $i+N$, although the following argument holds for open boundary conditions (OBC) as well. 

\subsection{$\pi$-magnon states}
It has been shown~\cite{Schecter2019} that scar states $\ket*{\mc{S}_n^{\mr{XY}}}$ can be constructed using the ferromagnetic state 
\begin{equation}
\ket*{\mc{S}^{\mr{XY}}_0}\coloneqq\bigotimes_{i\in\Lambda}\ket{-}_i  
\end{equation}
as a reference eigenstate and acting on it $n$ times with the creation operator
\begin{equation}
 Q^+_{\mr{XY}}\coloneqq\sum_{i\in\Lambda}(-1)^i\left(S_i^+\right)^2,
\end{equation}
that is,
\begin{equation}\label{eq:scar XY}
\ket*{\mc{S}_n^{\mr{XY}}}= \frac{1}{{\cal N}_n}(Q_{\mr{XY}}^+)^n\ket*{\mc{S}_0^{\mr{XY}}}.
\end{equation}
It is also known (and will be shown below) that $S_i^xS_{i+1}^x+S_i^yS_{i+1}^y$ annihilates all these states $\ket*{\mc{S}_n^{\mr{XY}}}$. The first term in Eq.~\eqref{eq:scar XY} is thus a sum of local annihilators and takes the role of $H_{\mr{ann}}$, while the single site terms (Zeeman field and single ion anisotropy) take the role of $H_{\rm spec}$. These scars are covered by the standard projector embedding.

To keep the paper self-contained we give here a simple proof of this statement by constructing the local projector embedding. We note, however, that these scar states have been  studied intensively, and we do not claim any novelty connected with this proof. 

Since the states Eq.~\eqref{eq:scar XY} are easily seen to be eigenstates of the second, longitudinal part in Eq.~\eqref{eq:spin-1 XY} 
it suffices to show that the individual XY terms locally annihilate them.

On the scar states \eqref{eq:scar XY} the possible local configurations of two sites $i, i+1$  are either $\ket*{\phi_-}\coloneqq\ket*{-,-},\,\ket*{\phi_0}\coloneqq(\ket*{+,-}-\ket*{-,+})/\sqrt{2}$, or $\ket*{\phi_+}\coloneqq\ket*{+,+}$. Thus we define a two-site projector as
\begin{equation}
    \begin{split}
        P^{\mr{XY}}_{i,i+1}&\coloneqq\left(\dyad*{\phi_-}+\dyad*{\phi_0}+\dyad*{\phi_+}\right)_{i,i+1}.
    \end{split}
\end{equation}
One can now easily check that the XY term associated with sites $i, i+1$ is annihilated by $P^{\mr{XY}}_{i,i+1}$, 
\begin{equation}\begin{split}
&\left(S_i^xS^i_{i+1}+S_i^yS^y_{i+1}\right)P_{i,i+1}^{\mr{XY}}\\&=P_{i,i+1}^{\mr{XY}}\left(S_i^xS^i_{i+1}+S_i^yS^y_{i+1}\right)=0.
\end{split}
\end{equation}
Indeed, the XY term annihilates $\ket*{\phi_\pm}$, which are configurations with  $S^z_{\mr{tot}}=\pm 2$.
Moreover, the action of the XY term restricted to the $S^z_{\mr{tot}}=0$ sector reads $\ket*{0,0}(\bra*{+,-}+\bra*{-,+})+h.c.$, which is seen to annihilate $\phi_0$. 

From the above, it follows that the XY model can be expressed in the projector-embedding form
\begin{equation}
    H_{\mr{XY}}=\sum_{i\in\Lambda}\left(1-P_{i,i+1}^{\mr{XY}}\right)h_i\left(1-P_{i,i+1}^{\mr{XY}}\right)+H_{\mr{spec}},
\end{equation}
where $h_i$ is a two-body operator acting on the sites $i,i+1$, and $H_{\mr{spec}}$ is the second term in Eq.~\eqref{eq:spin-1 XY}.

One can regard this as a special case of the formalism outlaid in Sec.~\ref{sec:main idea}, by identifying $\msf{P}$ with the identity operator (thus implementing the trivial embedding of the Hilbert space into  itself). 

\subsection{Bi-magnon state}
The above XY model hosts a second tower of scar states, the so-called bi-magnon states discovered by Schecter and Iadecola~\cite{Schecter2019} through numerical studies.
Chattopadhyay \textit{et.al.,}~\cite{Chattopadhyay2020} subsequently showed that they are indeed exact eigenstates, provided that the single ion anisotropy vanishes, $D=0$. However, they are not characterized by local annihilators and thus they constitute an interesting example for our Hilbert space extension formalism. 

The corresponding scar eigenstates were explicitly constructed~\cite{Chattopadhyay2020} and take the form
\begin{equation}\label{eq:bimagnon XY}
    \ket{\mc{B}^{\mr{XY}}_n}=\sum_{\{i_k\}_{k=1}^n\subset\Lambda}(-1)^{\sum_{k=1}^ni_k}\prod_{k=1}^nS_{i_k}^+S_{i_k+1}^+\bigotimes_{i\in\Lambda}\ket{-}_i,
\end{equation}
which suggests an interpretation in terms of bimagnon excitations on top of a ferromagnetic reference state. The sum is over all ordered $n$-tuples ${i_k}$ of mutually distinct sites ({this includes the possibility to have $i_{k+1}=i_k+1$). A proof that these are exact eigenstates will be sketched below.

We follow the procedure in Ref.~\cite{Chattopadhyay2020} to show that $\ket*{\mc{B}^{\mr{XY}}_n}$ is an eigenstate. To do so, we first fractionalize each spin-$1$ into two constituent spin-$1/2$, which are to be symmetrized in the end. Let $\Lambda_{\mr{frac}}\coloneqq\{1,\cdots,2N\}$ be the associated lattice of $2N$ spin-$1/2$. We define the map $F^{\mr{XY}}_i:\mb{C}^{2\otimes2}\rightarrow\mb{C}^3$ for the site $i$ as
\begin{equation}\label{eq:fractional spin XY}
    \begin{split}
        F^{\mr{XY}}_i&\coloneqq\ket*{+}_i\bra*{\uparrow\uparrow}_{2i-1,2i}+\ket*{-}_i\bra*{\downarrow\downarrow}_{2i-1,2i}\\&+\ket*{0}_i\left(\bra*{\uparrow\downarrow}+\bra*{\downarrow\uparrow}\right)_{2i-1,2i}.
    \end{split}
\end{equation}
A global map connecting the original spin-1 space and the (fractional) spin-1/2 is defined as $\msf{P}_{\mr{XY}}\coloneqq\prod_{i\in\Lambda}F_i^{\mr{XY}}$. Note that the map $F^{\mr{XY}}_i$ is not norm preserving, except in the subspace orthogonal to $S^z=0$. It is thus useful to express $F^{\mr{XY}}_i$ as a product of a norm-preserving map $F^{\mr{VBS}}_i$ and a subsequent map $G_i$ diagonal in the natural spin-1 basis as follows,
\begin{equation}\label{eq:local map vbs}
    \begin{split}
        F_i^{\mr{XY}}&=G_iF^{\mr{VBS}}_i,\\
        G_i&\coloneqq\left(\dyad*{+}+\sqrt{2}\dyad*{0}+\dyad*{-}\right)_i,\\
        F^{\mr{VBS}}_i&\coloneqq\ket*{+}_i\bra*{\uparrow\uparrow}_{2i-1,2i}+\ket*{-}_i\bra*{\downarrow\downarrow}_{2i-1,2i}\\&+\frac{1}{\sqrt{2}}\ket*{0}_i\left(\bra*{\uparrow\downarrow}+\bra*{\downarrow\uparrow}\right)_{2i-1,2i}.
    \end{split}
\end{equation}
The operator $F^{\mr{VBS}}_i$ implements the natural embedding of a spin-1 into the space of two fractionalized spins. We further define $G\coloneqq\prod_{i\in\Lambda}G_i$ and $\msf{F}_{\mr{VBS}}\coloneqq\prod_{i\in\Lambda}F^{\mr{VBS}}_i$ so that $\msf{P}_{\mr{XY}}$ is a product of $G$ and $\msf{F}_{\mr{VBS}}$ ($\msf{P}_{\mr{XY}}=G\msf{F}_{\mr{VBS}}$). With the map $\msf{P}_{\mr{XY}}$  the expression of the bi-magnon states simplifies to
\begin{equation}\label{eq:bi-magnon fractional}\begin{split}
    \ket*{\mc{B}^{\mr{XY}}_n}&=\frac{1}{\mc{N}_n}\msf{P}_{\mr{XY}}\ket*{\wtil{\mc{B}}^{\mr{XY}}_n}\\
    \ket*{\wtil{\mc{B}}^{\mr{XY}}_n}&\coloneqq\left(\sum_{i\in\Lambda}(-1)^i\dyad*{\uparrow\uparrow}{\downarrow\downarrow}_{2i,2i+1}\right)^n\bigotimes_{j\in\Lambda_{\mr{frac}}}\ket*{\downarrow}_j.
    \end{split}
\end{equation}

Let us now seek a corresponding Hamiltonian in the fractional spin-1/2 space. As $\msf{F}_{\mr{VBS}}$ has a non-trivial kernel, there exists more than one operator $\wtil{O}$ in the fractional spin-1/2 space that satisfy $O\msf{P}_{\mr{XY}}=\msf{P}_{\mr{XY}}\wtil{O}$. However, there is a natural choice for $\wtil{O}$. Indeed, for an arbitrary operator-valued function $f$, it holds that
\begin{equation}\label{eq:natural map}
    f(\{\bm{S}\}_i)\msf{F}_{\mr{VBS}}=\msf{F}_{\mr{VBS}}f(\{\bm{s}_{2i-1}+\bm{s}_{2i}\}_i),
\end{equation}
where $\bm{S}_i$ on the left-hand side (LHS) is a spin-1 operator, while $\bm{s}_j$ on the right-hand side (RHS) are spin-1/2 operators. Using Eq.~\eqref{eq:natural map} with the Hamiltonian in the place of $f(\{\bm{S}\}_i)$ defines a naturally lifted Hamiltonian in the fractional spin-1/2 space.
For operators in that extended space we again define an equivalence relation $\wtil{A}\sim_{\mr{XY}}\wtil{B}$ if $\msf{P}_{\mr{XY}}\wtil{A}=\msf{P}_{\mr{XY}}\wtil{B}$ is satisfied.

We now provide a brief sketch of the proof that $\ket*{\mc{B}^{\mr{XY}}_n}$ is an eigenstate of $H_{\mr{XY}}$, while details can be found in Appendix~\ref{sec:bi-magnon app}. The Zeeman term naturally translates to the fractional spin-1/2 space, 
\begin{eqnarray}\label{eq:Sz fractional}
    \left(\sum_{i\in\Lambda}S_i^z\right)\msf{P}_{\mr{XY}}&=&\msf{P}_{\mr{XY}}\left(\sum_{i\in\Lambda}\left(s_{2i-1}^z+s_{2i}^z\right)\right)\\&= &\msf{P}_{\mr{XY}} H_{\mr{spec}}
    ,
\end{eqnarray}
which yields the energies of the bi-magnon states. Indeed, we will now show that the XY term annihilates \textit{locally} the scar states in the fractional spin-1/2 space. From Eq.~\eqref{eq:bi-magnon fractional}, it is seen that local configurations of the scar states $\ket*{\wtil{\mc{B}}^{\mr{XY}}_n}$ at the sites $2i$ and $2i+1$ are either $\ket*{\uparrow\uparrow}$ or $\ket*{\downarrow\downarrow}$. We define the two-site projector onto the subspace spanned by those two states,
\begin{equation}
    P_{2i,2i+1}^{2\mr{-body}}\coloneqq\left(\dyad*{\uparrow\uparrow}+\dyad*{\downarrow\downarrow}\right)_{2i,2i+1}.
\end{equation}
Furthermore, the only possible local configurations at  four consecutive sites $2i\sim 2i+3$ are $\ket*{\phi_+}\coloneqq\ket*{\uparrow\uparrow\uparrow\uparrow},\ket*{\phi_0}\coloneqq(\ket*{\uparrow\uparrow\downarrow\downarrow}-\ket*{\downarrow\downarrow\uparrow\uparrow})/\sqrt{2}$, and $\ket*{\phi_-}\coloneqq\ket*{\downarrow\downarrow\downarrow\downarrow}$. Accordingly, we define the corresponding four-site projector,
\begin{equation}
    P^{4-\mr{body}}_{2i;2i+3}\coloneqq\left(\dyad*{\phi_+}+\dyad*{\phi_0}+\dyad*{\phi_-}\right)_{2i;2i+3}.
\end{equation}

\begin{widetext}
To lift the XY term in a suitable manner is more involved. In App.~\ref{sec:bi-magnon app} we show that it is equivalent to an expression     
\begin{equation}\begin{split}
    &\sum_{i\in\Lambda}\left(S_i^xS^x_{i+1}+S_i^yS^y_{i+1}\right)\msf{P}_{\mr{XY}} 
    =\msf{P}_{\mr{XY}}\left(\sum_{i\in\Lambda}\wtil{h}_{i,i+1}P^{2-\mr{body}}_{2i,2i+1}P^{2-\mr{body}}_{2i+2,2i+3}\right)+\msf{P}_{\mr{XY}}\left(\sum_{i\in\Lambda}\wtil{h}'_{i,i+1}\left(1-P^{2-\mr{body}}_{2i,2i+1}P^{2-\mr{body}}_{2i+2,2i+3}\right)\right),
    \end{split}
\end{equation}
\end{widetext}
The last term annihilates the scar states $\ket*{\wtil{\mc{B}}^{\mr{XY}}_n}$. 
Using the properties of the 2-body projectors and of $\msf{P}_{\mr{XY}}$,
one can further choose $\wtil{h}_{i,i+1}$ to be  equivalent to the following operator,
\begin{equation}\begin{split}
    \wtil{h}_{i,i+1}&\sim_{\mr{XY}}\left(\ket*{\downarrow\uparrow\uparrow\downarrow}+\ket*{\uparrow\downarrow\downarrow\uparrow}\right)\left(\bra*{\uparrow\uparrow\downarrow\downarrow}+\bra*{\downarrow\downarrow\uparrow\uparrow}\right)_{2i;2i+3}.
    \end{split}
\end{equation}
This operator annihilates the scar states, too, since its bra-part is orthogonal to all possible 4-site states, $\ket*{\phi_\pm}$ and $\ket*{\phi_0}$. It follows that the XY term annihilates any state that locally is a linear combination of $\ket*{\phi_\pm}$ and $\ket*{\phi_0}$.
This implies that the Hamiltonian lifted to the fractional spin-1/2 space can be expressed in local projector embedding form,
\begin{equation}\label{eq:xy fract pe}
    H_{\mr{XY}}\msf{P}_{\mr{XY}}=\msf{P}_{\mr{XY}}\left(\sum_{i\in\Lambda}\wtil{h}''_i\left(1-P^{4-\mr{body}}_{2i;2i+3}\right)+H_{\mr{spec}}\right),
\end{equation}
where $\wtil{h}''$ is a four-body operator and $H_{\mr{spec}}$ is the Zeeman term. The essence of this procedure is that the lifted Hamiltonian now consists of a sum of {\em local} annihilators, and that the scar space can be characterized as their common kernel. Equivalently, it is the space  left invariant by all 4-body projectors.

\section{Affleck-Kennedy-Lieb-Tasaki model}\label{sec:AKLT}
The second important example where the lifting to an enlarged Hilbert space allows to characterize the scar space via local annihilators, is the AKLT model. Its Hamiltonian is given by
\begin{equation}
    H_{\mr{AKLT}}=\sum_{i\in\Lambda}\left(\frac{1}{3}+\frac{1}{2}\left(\bm{S}_i\cdot\bm{S}_{i+1}\right)+\frac{1}{6}\left(\bm{S}_i\cdot\bm{S}_{i+1}\right)^2\right),
\end{equation}
where again $\Lambda=\{1,2,\cdots, N\}$ is a chain of sites with PBC, and $S^\alpha(\alpha=x,y,z)$ are the components of a spin-$1$ operator. Note that the Hamiltonian can be rewritten as a sum of projectors, i.e., $H_{\mr{AKLT}}=\sum_{i\in\Lambda}P_{i,i+1}^{S=2}$ where $P_{i,i+1}^{S=2}$ projects onto the  sector $S_{\mr{tot}}=2$ of the total spin of the two spins at sites $i$ and $i+1$.  

The scar states of the AKLT model $\ket*{\mc{S}_n^{\mr{AKLT}}}$ are obtained by applying a creation operator  $Q_{\mr{AKLT}}^+\coloneqq\sum_{i\in\Lambda}(-1)^i(S_i^+)^2$ to the AKLT ground state $\ket*{\Phi_{\mr{VBS}}}$~\cite{PhysRevB.98.235155},
\begin{equation}
    \ket*{\mc{S}_n^{\mr{AKLT}}}=\frac{1}{\mc{N}_n}\left(Q_{\mr{AKLT}^+}\right)^n\ket*{\Phi_{\mr{VBS}}}.
\end{equation}
$\ket*{\Phi_{\mr{VBS}}}$ is a valence-bond solid, i.e., a product of dimer singlets built from fractional spin-1/2 constituents,
\begin{equation}\begin{split}
    \ket*{\Phi_{\mr{VBS}}}&=\frac{1}{\mc{N}}\msf{F}_{\mr{VBS}}\bigotimes_{i\in\Lambda}\frac{1}{\sqrt{2}}\left(\ket*{\uparrow\downarrow}-\ket*{\downarrow\uparrow}\right)_{2i,2i+1}\\
    \equiv\frac{1}{\mc{N}}&\msf{F}_{\mr{VBS}}\bigotimes_{i\in\Lambda}\ket*{\Downarrow}_{2i,2i+1}\equiv\frac{1}{\mc{N}}\msf{F}_{\mr{VBS}}\ket*{\wtil{\Phi}_{\mr{VBS}}},
    \end{split}
\end{equation}
where we write a singlet $(\ket*{\uparrow\downarrow}-\ket*{\downarrow\uparrow})/\sqrt{2}$ as $\ket*{\Downarrow}$. We recall that {$\msf{F}_{\mr{VBS}}=\prod_{i\in\Lambda}F_i^{\mr{VBS}}$ is the map introduced in Eq.~\eqref{eq:fractional spin XY}}, which corresponds to the standard symmetrization of the two spin-1/2 with labels $2i-1$ and $2i$ to project on their spin-1 subspace.

Very similarly as for the bi-magnon scars of the XY model, we can express the scar states in the fractional spin-1/2 space as follows,
\begin{widetext}
\begin{equation}
\label{AKLT}
\begin{split}
    \ket*{\mc{S}_n^{\mr{AKLT}}}&=\frac{1}{\wtil{\mc{N}}_n}\msf{F}_{\mr{VBS}}\left(\sum_{i\in\Lambda}(-1)^i\dyad*{\Uparrow}{\Downarrow}_{2i,2i+1}\otimes\dyad*{\Uparrow}{\Downarrow}_{2i+2,2i+3}\right)^n\ket*{\wtil{\Phi}_{\mr{VBS}}}\equiv\frac{1}{\wtil{N}_n}\msf{F}_{\mr{VBS}}\ket*{\wtil{S}_n^{\mr{AKLT}}},
\end{split}
\end{equation}
\end{widetext}
where $\ket*{\Uparrow}\coloneqq\ket*{\uparrow\uparrow}$ is the triplet state with $S_{\mr{tot}}^z=1$. To understand why Eq.~\eqref{AKLT} holds, note that when two fractional spins at the sites $2i+1$ and $2i+2$ are excited by the term $(S_i^+)^2$ in $Q^+$, they must  previously have been in the down state. Thus, the neighboring fractional spins at the sites $2i-1$ and $2i+3$ must have been (and still remain) in the up state because of the singlet structure in $\ket*{\wtil{\Phi}_{\mr{VBS}}}$. This explains the shape of the operator appearing in Eq.~\eqref{AKLT}.

To lift the AKLT Hamiltonian, we first introduce the operators $P_{i,j}^{(S,M)}$ that project two (non-fractionalized) spins-1 on the sites $i$ and $j$ onto the sector of total spin $S$ and $S^z=M$. One finds
\begin{equation}\begin{split}
    P_{i,i+1}^{(S,M)}\msf{F}_{\mr{VBS}}&=\msf{F}_{\mr{VBS}}P\bigg[\left(\sum_{j=2i-1}^{2i+2}s_j\right)^2=S(S+1), \sum_{j=2i-1}^{2i+2}s_j^z=M\bigg]\\
    &\equiv\msf{F}_{\mr{VBS}}\wtil{P}_{i,i+1}^{(S,M)},
    \end{split}
\end{equation}
where the projector on the RHS, denoted $\wtil{P}_{i,i+1}^{(S,M)}$, projects the four spin-1/2 on the sites $2i-1\sim 2i+3$ onto the sector with total spin $S$ and $S^z=M$. 

With these operators we can now obtain a natural lifted Hamiltonian in the fractional spin-1/2 space, $H_{\mr{AKLT}}\msf{F}_{\mr{VBS}}=\msf{F}_{\mr{VBS}} \sum_{M=-2}^2 \wtil{P}_{i,i+1}^{(2,M)} \equiv\msf{F}_{\mr{VBS}}\wtil{H}_{\mr{AKLT}}$, or reorganized differently in the form  
\begin{equation}\label{eq:aklt frac}
    \begin{split}
        \wtil{H}_{\mr{AKLT}}&=\wtil{H}_{\mr{spec}}+\wtil{H}^<_{\mr{ann}}+\wtil{H}^>_{\mr{ann}},
    \end{split}
\end{equation}
where 
\begin{equation}\label{eq:aklt decomp}
    \begin{split}
        \wtil{H}_{\mr{spec}}&\coloneqq\sum_{i\in\Lambda_{\mr{frac}}}s_i^z\\
        \wtil{H}_{\mr{ann}}^<&\coloneqq\sum_{i\in\Lambda}\sum_{M\leq0}\wtil{P}^{(2,M)}_{i,i+1}\\
        \wtil{H}_{\mr{ann}}^>&\coloneqq\sum_{i\in\Lambda}\left(\wtil{P}^{(2,1)}_{i,i+1}+\wtil{P}^{(2,2)}_{i,i+1}-s_{2i}^z-s_{2i+1}^z\right),
    \end{split}
\end{equation}
$\Lambda_{\mr{frac}}=\{1,\cdots,2N\}$ being the lattice of fractional spins. As before $\wtil{H}_{\mr{AKLT}}$ is determined only up to embedding equivalence of operators, ($\wtil{A} \sim_{\mr{VBS}} \wtil{B}$ if $\msf{F}_{\mr{VBS}}\wtil{A}=\msf{F}_{\mr{VBS}}\wtil{B}$). We now look for an equivalent operator  $\sim_{\mr{VBS}} \wtil{H}_{\mr{AKLT}}$ that takes the generalized projector-embedding form Eq.~\eqref{eq:generalized pe} (see Appendix.~\ref{sec:aklt app} for a detailed derivation).

The scar states in the spin-1/2 space $\ket*{\wtil{S}_n^{\mr{AKLT}}}$ in Eq.~\eqref{AKLT} are easily seen to be eigenstates of $\wtil{H}_{\mr{spec}}$. They are also locally annihilated by each term in $\wtil{H}_{\mr{ann}}^<$. Indeed those annihilate the ground state as well as those components of excited scar states that haven't been acted on {by a spin raising operator} in that location. 
 On the other hand, a local excitation affecting two selected neighboring sites in ($\ket*{\wtil{\mc{S}}_n^{\mr{AKLT}}}$) may lead to a total spin 2 only if at least one of them has been acted upon by a spin raising operator with $\Delta M=2$. But the ground state configuration of the two sites, has at least $M\geq-1$ (as it contains no $S=2$ component) and thus local excitations on such site pairs must have $M>0$.  

It remains to find an operator $\wtil{H}'_{\mr{ann}}$ that is a sum of local terms that annihilate the scar states, and is equivalent to $\wtil{H}^>_{\mr{ann}}$ ($\wtil{H}'_{\mr{ann}}\sim_{\mr{VBS}}\wtil{H}^>_{\mr{ann}}$).

To do so, we consider possible local configurations of the scar states. From Eq.~\eqref{AKLT} it is clear that the local configurations of pairs of fractional spin sites ($2i$,$2i+1$) are either $\ket*{\Downarrow}$ or $\ket*{\Uparrow}$. We thus define the following two-body projector,
\begin{equation}
    P_{2i,2i+1}^{2\mr{-body}}\coloneqq \dyad*{\Uparrow}_{2i,2i+1}+\dyad*{\Downarrow}_{2i,2i+1}.
\end{equation}
One can also easily see that a 6-site configuration $\ket*{\Downarrow,\Uparrow,\Downarrow}$ cannot occur since the creation operator $Q^+_{\mr{AKLT}}$ always excites a pair of ``magnons'' $\ket*{\Uparrow}$ next to each other. 
Furthermore, on six consecutive sites $2i-2\sim2i+3$, only the following states can occur:
\begin{equation}\label{eq:AKLT scar local}
    \begin{split}
        \ket*{\phi^{\Vec{\sigma}}_0}&\coloneqq\ket*{\sigma_1,\sigma_2,\Downarrow,\sigma_3,\sigma_4}_{2i-2;2i+3}\\
        \ket*{\phi_1}&\coloneqq\frac{1}{\sqrt{2}}\big(\ket*{\Uparrow,\Uparrow,\Downarrow}-\ket*{\Downarrow,\Uparrow,\Uparrow}\big)_{2i-2;2i+3}\\
        \ket*{\phi_2}&\coloneqq\ket*{\Uparrow,\Uparrow,\Uparrow}_{2i-2;2i+3},
    \end{split}
\end{equation}
where $\Vec{\sigma}=(\sigma_1,\sigma_2,\sigma_3,\sigma_4)\,\,(\sigma_i=\uparrow\downarrow)$ is a four-spin configuration. 
This is confirmed by distinguishing the cases that (i) the central pair has not been excited, (ii) it has been excited with one of its neighbor pairs while the other pair was not excited, or (iii) all pairs have been excited.
We denote the projector on the space spanned by  these possible states by $P^{6\mr{-body}}_{2i-2;2i+3}(=\sum_{\Vec{\sigma}}\dyad*{\phi_0^{\Vec{\sigma}}}+\dyad*{\phi_1}+\dyad*{\phi_2})$.

We now split the action of $\wtil{H}^>_{\mr{ann}}$ into two pieces, acting on this subspace and its complement, respectively.
After some calculations (see Appendix.~\ref{sec:aklt app}) one finds that  $\wtil{H}^>_{\mr{ann}}$ is equivalent to,
\begin{equation}\label{eq:aklt annihilate pre}
    \wtil{H}^>_{\mr{ann}}\sim_{\mr{VBS}}\sum_{i\in\Lambda}h_i
    +h'_i\left(1-P^{6-\mr{body}}_{2i-2;2i+3}\right),
\end{equation}
where $h'_i$ is some operator and 
\begin{equation}\begin{split}
    h_i&\coloneqq-\frac{1}{2}\ket*{\psi}\left(\bra*{\Downarrow,\Uparrow,\Uparrow}+\bra*{\Uparrow,\Uparrow,\Downarrow}\right)_{2i-2;2i+3}\\
    \ket*{\psi}&\coloneqq\frac{1}{\sqrt{2}}\left(\ket*{\uparrow\downarrow,\Uparrow,\Uparrow}-\ket*{\Uparrow,\Uparrow,\downarrow\uparrow}\right).
    \end{split}
\end{equation}
The second term in Eq.~\eqref{eq:aklt annihilate pre} annihilates the scar states as it projects out all the local configurations that may occur. 
The first term also does so, because the bra-part of $h_i$ is orthogonal to all $\ket*{\phi_i}\,\,(i=0,1,2)$. For this reason we can write
\begin{eqnarray}\label{eq:aklt annihilate}
    \wtil{H}^<_{\mr{ann}}\sim_{\mr{VBS}}\sum_{i\in\Lambda}h_i+h'_i\left(1-P^{6-\mr{body}}_{2i-2;2i+3}\right)\\
    =\sum_{i\in\Lambda}(h_i+h'_i)\left(1-P^{6-\mr{body}}_{2i-2;2i+3}\right).
\end{eqnarray}

Putting everything together, we have obtained a form for the lifted Hamiltonian which indeed takes a {\em local} projector-embedding form,
\begin{equation}\label{eq:aklt projector embedding}\begin{split}
    &H_{\mr{AKLT}}\msf{F}_{\mr{VBS}}\\&=\msf{F}_{\mr{VBS}}\left(\wtil{H}_{\mr{spec}}+\wtil{H}^<_{\mr{ann}}+\sum_{i\in\Lambda}(h_i+h'_i)\left(1-P^{6\mr{-body}}_{2i-2;2i+3}\right)\right),
    \end{split}
\end{equation}
where $h_i+h'_i$ is a local 6-body operator.

\section{Spin-1/2 domain-wall preserving model}\label{sec:dw model}

{We have seen that the modified PXP model, the spin-1 XY model, and the AKLT model all share the structure of a hidden projector embedding. That is, there exists an enlarged Hilbert space, and a map $\msf{P}$ that connects it to the original Hilbert space, satisfying the relation $H\msf{P}=\msf{P}(\wtil{H}_{\mr{spec}}+\sum_i\wtil{h}_i\wtil{P}_i)$. Parent scar states $\ket*{\wtil{\mc{S}}_n}$, defined in the enlarged space, are eigenstates of $\wtil{H}_{\mr{spec}}$ and are annihilated by $\wtil{P}_i$. The corresponding scar states in the original space are simply obtained by projection as  $\msf{P}\ket*{\wtil{\mc{S}}_n}$.}

{Let us now review in this context the domain wall preserving spin-1/2 model of Eq.~\eqref{eq:Hdw}, as defined and discussed in Ref.~\cite{Iadecola2020DW}. This model can be cast in projector embedding form directly in the original spin-1/2 Hilbert space, and thus an extension of the Hilbert space is not really needed. However, we may use the fact that the model preserves the number of domain walls. The corresponding subspace can be interpreted as a constrained space, to which the formalism developed above can be applied again.
This framework allows to 
reveal a simple structure of the various sets of scar states in this model.}

A key observation is that the tower of scar states, given explicitly below as 
``pyramidal scars''  $\ket*{\mc{S}^{\mr{Pyr}}_{n,m}}$ (Eq.~\eqref{eq:pyramid}), can be written as projected states 
$P^{\mr{dw}}_n\ket*{\mc{S}^{\mr{pre}}_{n+m}}$, where 
$P^{\mr{dw}}_n$ is a domain wall number projector that will be defined below. The parent state $\ket*{\mc{S}^{\mr{pre}}_{n+m}}$ has a simple structure that allows local annihilators to be identified easily. 



\subsection{Model and scar states}\label{sec:model and scar dw}
The spin-$1/2$ domain-wall preserving model {is defined as,}
\begin{equation}\label{eq:Hdw}\begin{split}
    H_{\mr{DW}}&=\sum_{i\in\Lambda}\lambda\left(\sigma^x_i-\sigma^z_{i-1}\sigma^x_i\sigma^z_{i+1}\right)+\Delta\sigma^z_i+J\sigma^z_i\sigma^z_{i+1}\\
    &{\equiv H_\lambda+H_{\mr{Zeeman}}+H_{\mr{Ising}},}
    \end{split}
\end{equation}
where $\Lambda\coloneqq\{1,2,\cdots,N\}$. 

We assume periodic boundary conditions and an even number of sites $N$. We denote the first term by $H_\lambda$ following the convention in Ref.~\cite{Mark2020}. This model conserves the number of domain walls, i.e., interfaces between up- and down-spins. The total number of the domain walls is counted by the operator $N_{\mr{dw}}\coloneqq\sum_{i\in\Lambda}(\dyad*{\downarrow\uparrow}+\dyad*{\uparrow\downarrow})_{i,i+1}$, which commutes separately with $H_\lambda$, $H_{\mr{Zeeman}}$ and $H_{\mr{Ising}}$. With periodic boundary conditions $N_{\mr{dw}}$ is always even.
Let us then define the projector $P^{\mr{dw}}_n$ onto the sector of $2n$ domain walls, which commutes with $H_{\mr{DW}}$. Note that $P^{\mr{dw}}_n$ is a global operator which cannot be expressed as a product of local operators.

The ``domain wall'' scar states which we denote by $\ket*{\mc{S}_n^{\mr{DW}}}$ are constructed from the ferromagnetic state
\begin{equation}
    \ket*{\mc{S}_0^{\mr{DW}}}\coloneqq\bigotimes_{i\in\Lambda}\ket*{\downarrow}_i,
\end{equation}
by applying the creation operator $Q^+_{\mr{DW}}$ to it,
\begin{equation}\label{eq:|Sn> dw}\begin{split}
    \ket*{\mc{S}_n^{\mr{DW}}}&=\frac{1}{\mc{N}_n}(Q_{\mr{DW}}^+)^n\ket*{\mc{S}_0^{\mr{DW}}},\\
    Q_{\mr{DW}}^+&\coloneqq\sum_{i\in\Lambda}(-1)^iP_{i-1}\sigma_i^+P_{i+1}.
    \end{split}
\end{equation}
where $P_i=\dyad*{\downarrow}_i$. Note that these states never contain two adjacent up-spins, and thus obey a Rydberg-like constraint.


Let us now show that $\ket*{\mc{S}_n^{\mr{DW}}}$ is an exact eigenstate of $H_{\mr{DW}}$ by casting the latter into projector-embedding form. We first show that $H_\lambda$ annihilates the scar states. To do so, we rewrite $H_\lambda$ as 
\begin{equation}\label{eq:local ann DW}\begin{split}
    H_\lambda&=2\lambda\sum_{i\in\Lambda}\left(P_{i-1}\sigma_i^xn_{i+1}+n_{i-1}\sigma_i^xP_{i+1}\right)\\
    &=2\lambda\sum_{i\in\Lambda}\ket*{\downarrow\uparrow\uparrow\downarrow}\left(\bra*{\downarrow\uparrow\downarrow\downarrow}+\bra*{\downarrow\downarrow\uparrow\downarrow}\right)_{i-1;i+2}\\
    &+2\lambda\sum_{i\in\Lambda}\left(\dyad*{\downarrow\uparrow\uparrow\uparrow}{\downarrow\downarrow\uparrow\uparrow}+\dyad*{\uparrow\uparrow\uparrow\downarrow}{\uparrow\uparrow\downarrow\downarrow}\right)_{i-1;i+2}\\
    &+2\lambda\sum_{i\in\Lambda}\left(\dyad*{\downarrow\downarrow\uparrow}{\downarrow\uparrow\uparrow}+\dyad*{\uparrow\downarrow\downarrow}{\uparrow\uparrow\downarrow}\right)_{i-1,i,i+1},
    \end{split}
\end{equation}
where $n\coloneqq\dyad{\uparrow}$. As the bra part of the operators in the last two terms 
contains two consecutive up-spins $\bra*{\uparrow\uparrow}$, they annihilate the scar states which obey the Rydberg constraint. The first of the three four-body terms annihilates the scar states, too, as in the latter configurations isolated up-spins come with site-alternating signs, which is orthogonal to the bra part of the first term, $\ket*{\downarrow\uparrow\downarrow\downarrow}+\ket*{\downarrow\downarrow\uparrow\downarrow}$. We denote the projector on that four-site configuration by $P^{4-\mr{body}}_{i-1;i+2}$. 
From the above it follows that $H_\lambda$ annihilates the scar states $\ket*{{\mc{S}}^{\mr{DW}}_n}$ of Eq.~(\ref{eq:|Sn> dw}).


The scar states are trivially eigenstates of $H_{\mr{spec}}$,  the sum of the remaining Zeeman term and the Ising interaction in Eq.~(\ref{eq:Hdw}). Therefore, the Hamiltonian can be cast into the projector-embedding form, 
\begin{equation}\label{eq:sm dw}
    H_{\mr{DW}}=H_{\mr{spec}}+\sum_{i\in\Lambda}\left(h_iP_{i-1;i+2}^{4-\mr{body}}+h'_in_in_{i+1}\right),
\end{equation}
{where $h_i$ and $h'_i$ are some local four-body operators that multiply the scar annihilators $P_{i-1;i+2}^{4-\mr{body}}$ and $n_in_{i+1}$ from the left.

\subsection{Tower of pyramid scar states}
In Ref.~\cite{Mark2020}, the so-called pyramid states were found to constitute further sets of exact scar eigenstates of $H_{\mr{DW}}$. These towers are constructed from the domain wall scar states $\ket*{\mc{S}_n^{\mr{DW}}}$ by further acting with a non-local operator $Q^+_{\mr{Pyr}}$,
\begin{equation}\label{eq:pyramid}
    \begin{split}
        \ket*{\mc{S}_{n,m}^{\mr{Pyr}}}&=\frac{1}{\mc{N}_{n,m}}\left(Q_{\mr{Pyr}}^+\right)^m\ket*{\mc{S}^{\mr{DW}}_n},\\
        {Q_{\mr{Pyr}}^+}&={\sum_{i\in\Lambda}\sum_{l=1}^{N-2}\prod_{k=1}^ln_{i-k}\sigma_i^+P_{i+1}.}
    \end{split}
\end{equation}
{This raising operator picks out ferromagnetic domains of up-spins and extends them by one more spin on their right edge, provided that this does not merge the domain with the subsequent one.}
{The pyramidal states can thus also be written in the  form} 
\begin{equation}\label{eq:S pyr}
    {\ket*{\mc{S}^{\mr{Pyr}}_{n,m}}=\frac{1}{\mc{N}_{n,m}}\sum_{i_1<\cdots<i_n}\sum_{\{l_k\}_{k=1}^n}(-1)^{\sum_ki_k}\ket*{\{l_k\},\{i_k\}},}
\end{equation}
{where $i_k\,(1\leq k\leq n)$ is the left-most site of the $k$-th ferromagnetic domain of up-spins, $l_k$ being its length. Thus the state $\ket*{\{l_k\},\{i_k\}}$ should be read as}
\begin{equation}
    \ket*{\{l_k\},\{i_k\}}=\ket*{\downarrow\cdots\downarrow\underbrace{\uparrow\cdots\uparrow}_{i_1\sim i_1+l_1-1}\downarrow\cdots\downarrow\underbrace{\uparrow\cdots\uparrow}_{i_n\sim i_n+l_n-1}\downarrow\cdots},
\end{equation}
where $i_1\sim i_1+l_1-1$ indicates that a ferromagnetic domain of length $l_1$ occupies the sites $i_1\sim i_1+l_1-1$. While $n$ is the number of ferromagnetic domains, {$m$ is determined by the total number of up-spins minus the number of domains, $m=\sum_kl_k-n$.}

In Ref.~\cite{Mark2020}, Mark \textit{et al.,} showed that $\ket*{\mc{S}_{n,m}^{\mr{Pyr}}}$  are exact eigenstates by directly solving the stationary Schr\"odinger equation. Here we provide a simpler proof 
based on the projector-embedding.
This consists of two steps. First, we define  ``pre-pyramid'' states $\ket*{\mc{S}^{\mr{pre}}_n}$ and show that they are exact eigenstates in the absence of the Ising interaction. 
In the second step we show that the projection of the pre-pyramid states onto a sector of fixed domain wall number yields the desired pyramid states. Those remain exact eigenstates even in the presence of the Ising interaction.

\subsubsection{{Pre-pyramid states}}
Consider the ``pre-pyramid'' states
\begin{equation}\label{eq:pre-pyr}\begin{split}
    {\ket*{\mc{S}^{\mr{pre}}_n}}&{=\frac{1}{\mc{N}_n}U_{\mr{dw}}\left(\sum_{i\in\Lambda}\sigma_i^+\right)^n\bigotimes_{j\in\Lambda}\ket*{\downarrow}_j}\\
    &{=\frac{1}{\mc{N}'_n}U_{\mr{dw}}\left(\sum_{i\in\Lambda}\sigma_i^-\right)^{N-n}\bigotimes_{j\in\Lambda}\ket*{\uparrow}_j,}
    \end{split}
\end{equation}
{where $U_{\mr{dw}}\coloneqq\exp[i\pi\sum_{j\in\Lambda}j\dyad*{\downarrow\uparrow}_j]$ is a unitary transformation which gives momentum $\pi$ to each ferromagnetic domain.}

These states are superpositions of configurations with equal magnetization and thus are eigenstates of the Zeeman term. Later we will project them to  subspaces of constant domain wall number, which turns them to eigenstates of the  Ising interaction as well.  
The Zeeman and Ising terms will constitute $H_{\mr{spec}}$ in the final projector embedding.

We next show that the remaining $H_\lambda$ can be expressed as a sum of \textit{quasi-}local annihilators. To this end we split $H_\lambda$ as $H_\lambda=H^+_\lambda+H^-_\lambda$ where $H^\pm_\lambda=2\lambda\sum_i (P_{i-1}\sigma_i^\pm n_{i+1}+n_{i-1}\sigma_i^\pm P_{i+1})$. One finds that $H^\pm_\lambda$ can be rewritten as
\begin{widetext}
\begin{equation}\label{eq:H_lam pm}\begin{split}
    H^+_\lambda&
    =2\lambda\sum_{i\in\Lambda}\bigg[\sum_{k=2}^{N-2}\ket*{\downarrow\underset{k\,\text{sites}}{\underline{\uparrow\cdots\uparrow}}\downarrow}\bigg(\bra*{\downarrow\downarrow\underset{k-1\,\text{sites}}{\underline{\uparrow\cdots\uparrow}}\downarrow}+\bra*{\downarrow\underset{k-1\,\text{sites}}{\underline{\uparrow\cdots\uparrow}}\downarrow\downarrow}\bigg)_{i-1;i+k}
    +\ket*{\downarrow\underset{N-1\,\text{sites}}{\underline{\uparrow\dots\uparrow}}}\bigg(\bra*{\downarrow\downarrow\underset{N-2\,\text{sites}}{\underline{\uparrow\cdots\uparrow}}}+\bra*{\downarrow\underset{N-2\,\text{sites}}{\underline{\uparrow\cdots\uparrow}}\downarrow}\bigg)_{i-1;i+N-2}\bigg]\\
    H^-_\lambda&
    =2\lambda\sum_{i\in\Lambda}\bigg[\sum_{k=2}^{N-2}\ket*{\uparrow\underset{k\,\text{sites}}{\underline{\downarrow\cdots\downarrow}}\uparrow}\bigg(\bra*{\uparrow\uparrow\underset{k-1\,\text{sites}}{\underline{\downarrow\cdots\downarrow}}\uparrow}+\bra*{\uparrow\underset{k-1\,\text{sites}}{\underline{\downarrow\cdots\downarrow}}\uparrow\uparrow}\bigg)_{i-1;i+k}
    +\ket*{\uparrow\underset{N-1\,\text{sites}}{\underline{\downarrow\cdots\downarrow}}}\bigg(\bra*{\uparrow\uparrow\underset{N-2\,\text{sites}}{\underline{\downarrow\cdots\downarrow}}}+\bra*{\uparrow\underset{N-1\,\text{sites}}{\underline{\downarrow\cdots\downarrow}}\uparrow}\bigg)_{i-1;i+N-2}\bigg].
    \end{split}
\end{equation}
\end{widetext}
Each of these terms annihilates the pre-pyramid states. Indeed they all   act non-trivially only on a symmetric superposition of two local configurations where two down spins frame a ferromagnetic up-island plus one further down spin (or vice versa, with up and down interchanged), while the pre-pyramid states only contain the antisymmetric superposition of such local configurations.


{It follows that the pre-pyramid states are exact eigenstates of the Hamiltonian except for the Ising interaction. Excluding the latter for the moment, we thus have the projector-embedding form $H_\lambda+H_{\mr{Zeeman}}=H_{\mr{spec}}+\sum_ih_iP_i$, where $H_{\mr{spec}}$ is the Zeeman term, and the $P_i$ are projectors on symmetric superpositions of  configurations as described above.

{Note that these  annihilating projectors are \textit{quasi-}local. While each one has finite support, the diameter of the support is not uniformly bounded over the set of all annihilators. 
In contrast to the spin-1 XY and the AKLT model, we have not found any map which would make the annihilators truly local in a larger space. Nevertheless, we emphasize that the annihilator $H_\lambda$ is not an ``as-a-whole'' annihilator since we can decompose it explicitly as a sum of an \textit{extensive} number of annihilators, many of which are local and do not span a finite fraction of the system. Those that do affect an extensive number of sites have instead a very low rank. For the characterization of scar states with a finite density of domain walls they play a negligible role.}}

\subsubsection{{Projection to sectors of constant domain wall number}}

{The pre-pyramid state is a superposition of states with different numbers of domain walls. However, the Hamiltonian conserves the number of domain walls. We may thus project the pre-pyramid state onto the sector with a fixed number of $2m$ domain walls,  $P_m^{\mr{dw}}\ket*{\mc{S}^{\mr{pre}}_n}$. Since $P^{\mr{dw}}_m$ commutes with $H_\lambda+H_{\mr{Zeeman}}$, the projected state $P_m^{\mr{dw}}\ket*{\mc{S}^{\mr{pre}}_n}$ is still an eigenstate. Furthermore, it is also an eigenstate of the Ising interaction, which indeed reduces to a constant in every sector with a fixed number of domain walls.}

Let us now show that these states are equivalent to the pyramidal states of Eq.~\eqref{eq:S pyr}. To this end we change the labelling to $n \to (n'+m')$, $m\to n'$
and expand the projections of the states \eqref{eq:pre-pyr} in terms of the positions and lengths of the $n'$ up-domains as

\begin{equation}\label{eq:Spre project}
    P^{\mr{dw}}_{n'}\ket*{\mc{S}^{\mr{pre}}_{n'+m'}}=\frac{1}{\mc{N}_{n',m'}}\sum_{i_1<\cdots<i_{n'}}\sum_{\{l_k\}_{k=1}^{n'}}(-1)^{\sum_{k=1}^{n'} i_k}\ket*{\{l_k\},\{i_k\}},
\end{equation}
{which is indeed identical to Eq.~\eqref{eq:S pyr}.}

From the above the following structure emerges: Since the Hamiltonian conserves the domain wall number, we can consider the subspaces of fixed domain wall number separately as constrained Hilbert spaces. Each of them turns out to host a tower of scar states, which can be obtained as a projection (with the corresponding $P^{\mr{dw}}_{n'}$) from the parent scar states of Eq.~\eqref{eq:pre-pyr}. Those in turn have a simple characterization via the projector embedding form of the Hamiltonian (excluding the Ising term which simply embodies a linear coupling to the domain wall constraint and thus is constant on the considered subspaces). 
}

Here the domain wall projector  $P^{\mr{dw}}_n$ takes the role of the embedding map $\msf{F}$, similarly as $\msf{F}_{\mr{VBS}}$ discussed above. However, unlike in the previously  discussed cases, the projector $P^{\mr{dw}}_n$ cannot be expressed as a product of local operators. This is because the domain wall number is a global constraint that concerns the system as a whole. 

In the next section, we will discuss the resemblance of this non-local projection and the scar states to similar structures that appear in topological phases of gauge theories.


\section{Similarities with gauge theories}\label{sec:lgt}
As we have seen in the previous sections,  non-trivial quantum many-body scar states often have the structure $\msf{P}\ket*{\wtil{\mc{S}}_n}$ where $\msf{P}$ projects a parent state $\ket*{\wtil{\mc{S}}_n}$ onto the ``physical'' subspace, such as the $S=1$ Hilbert space for the AKLT model.  $\ket*{\wtil{\mc{S}}_n}$ often has a relatively simple structure, with low entanglement in the enlarged Hilbert space. Local degrees of freedom in the enlarged space, such as the $S=1/2$ spins in AKLT, may then sometimes be interpreted as fractionalized entities.

Such a fractionalization is ubiquitous in topological phases of matter as well. Thus it is natural to ask whether there is any connection between the scar states we have discussed and wavefunctions in many body systems with topological order. 

In this section, we point out structural similarities between ground state wave functions in the deconfined phase of gauge theories and some of the scar states we have discussed. 

Gauge theories arise in the context of systems that obey local constraints between their degrees of freedom, such as a Gauss' law. The constraints are enforced by projectors $\msf{G}_{\mr{gauge}}$, that project onto the physical subspace $\mc{V}_{\mr{gauge}}$, which obeys the constraint or Gauss' law,  $\msf{G}_{\mr{gauge}}\ket*{\psi}=+\ket*{\psi}$ for $\psi \in \mc{V}_{\mr{gauge}}$. This projector $\msf{G}_{\mr{gauge}}$ can be viewed as analogous to the embedding map $\msf{P}$ for scar states that descend from a parent state in an enlarged, unconstrained Hilbert space.
Ground states of gauge theories in deconfined phases are usually superpositions of all constraint-obeying configurations in a given topological sector. 
However, they can often be expressed as the projection of a low entangled state in the enlarged Hilbert space where  Gauss' law is not imposed. 
For every physical state, one has the gauge freedom to choose a particular parent state  that projects to it. Choosing such a parent state amounts to gauge fixing. This can be done, e.g., by modifying the Hamiltonian in the enlarged Hilbert space, such that its restriction to the physical space does not change, but such that it selects the desired, simple parent state as the ground state in the enlarged space. This is very much analogous to our using the freedom to choose $\wtil{H}$ such that it admits a particularly simple scar subspace (resulting in a simple projector embedding form). 

In gauge theories there often exist non-local (string-like) projectors $P_{\mr{top}}$ that commute with the Hamiltonian and define distinct topological sectors. Such constraint enforcing projectors are analogous to the $P_n^{\rm dw}$ that project a parent scar state onto the scar states within the sectors of fixed domain wall number.
From this viewpoint the multiplicity of the towers of pyramidal states is  an analogue of the topological degeneracy in the gauge theories.

In App.~\ref{app:gauge}, we describe in more detail two explicit examples, Ising gauge theory (IGT) and the quantum dimer model (QDM), the construction of their ground states and the similarity with hidden projector embeddings of QMBS.

\section{Conclusion}
\label{sec:conclusion}
In this paper we have scrutinized the generality of the   local projector embedding paradigm, where the Hamiltonian can be 
cast into the from $H=H_{\mr{spec}}+H_{\mr{ann}}$ and $H_{\mr{ann}}$ is a sum of local projectors which annihilate the scar states.
We have shown that several scar-hosting models that were previously thought not to be covered by this paradigm can in fact be understood within one and the same framework, provided it is slightly generalized. Our finding thus reveals the hidden structure of these models and provides a unified understanding of quantum many-body scar states.

For the bi-magnon scar states of the spin-1 XY model and the AKLT model, we have defined an enlarged Hilbert space where the fundamental degrees of freedom ($S=1$) are fractionalized, together with a map $\msf{P}$ that projects down to the original, physical Hilbert space. Defining a suitable lifted Hamiltonian $\wtil{H}$ in the enlarged Hilbert space, i.e., $H\msf{P}=\msf{P}\wtil{H}$, it can be cast in local projector embedding form. The lifted scar subspace is obtained as the non-trivial intersection of the kernels of all local annihilators. Simultaneously that subspace is found to be a union of eigenspaces of the remaining spectrum generating term $H_{\mr{spec}}$ in the Hamiltonian. 

We have discussed the similarity of our construction to find scars in models with constraints to  features that arise when constructing wavefunctions in the deconfined phase of lattice gauge theories.
There  Gauss' law takes the role of the constraint, and is imposed by a projector analogous to the above map $\msf{P}$. 
Deconfined ground states may look very complex in the physical space where Gauss' law is imposed, while they often take a simple form, such as a product state, when lifted to the unconstrained Hilbert space, upon choosing a proper gauge. 
The analogue in the scar problem is the fact that we found a simple projector embedding form of the Hamiltonian only in the enlarged, unconstrained Hilbert spaces, and upon properly choosing a suitable parent Hamiltonian.
Both in lattice gauge theories and in scar models, the projection onto the physical subspace transforms the simple parent states into entangled states which are more complex superpositions of configurations that obey the respective constraints.


It may be useful to apply our formalism to study the ground state and excitations (and possibly approximate scar states) in the $\mb{Z}_2$ spin liquid state  proposed to occur in Rydberg atom arrays, considering that the long-lasting oscillations ensuing from an initial N\'eel state in the PXP model have sometimes been interpreted within the framework of lattice gauge theories (e.g., Ref.~\cite{PhysRevX.10.021041}). 

Given that our approach can be suitably generalized to a (modified) PXP model~\cite{PhysRevA.107.023318}, it would  be interesting to extend it to further models that appear to host (approximate) scars  {whose origin is, however, yet unclear, such as 
e.g., those introduced in Refs.~\cite{Bull2019,Hudomal2020}. 

Finally it would be worthwhile to establish a precise relation between our approach of local annihilators on an extended space and the commutant algebra discussed in Ref.~\cite{https://doi.org/10.48550/arxiv.2209.03377}. Especially, it would be interesting to understand whether and how the latter can be generalized to Hilbert space embeddings as we discussed here. }

\acknowledgements
We thank M. Sigrist and K. Pakrouvski for their useful comments and discussions. This work was supported by grant No. 
 204801021001 of the Swiss National Science Foundation.
\normalem
\bibliographystyle{unsrt}
\bibliography{reference}
\appendix
\section{Projector-embedding construction of bi-magnon scar states}
\label{sec:bi-magnon app}

{In the main text, we sketched that the bi-magnon scars $\ket*{\mc{B}^{\mr{XY}}_n}$ defined in Eqs.~(\ref{eq:bimagnon XY},\ref{eq:bi-magnon fractional}) can be understood in terms of a  local projector-embedding on a suitably extended Hilbert space.} In this appendix, we elaborate this point by constructing a parent Hamiltonian that has the sought projector embedding structure, with the claimed parent scar states. At the same time  this furnishes a detailed proof that the projected scar states are indeed exact eigenstates of $H_{\mr{XY}}$, provided that the single ion anisotropy vanishes, $D=0$.

\subsection{Lifting of the Hamiltonian}
Like the PXP model and the AKLT model, as the mapping from fractionalized spins to $S=1$ spins $\msf{P}_{\mr{XY}}$ is not injective, there is no unique choice for the operator $\wtil{O}$ in the fractional spin space that satisfies $O\msf{P}_{\mr{XY}}=\msf{P}_{\mr{XY}}\wtil{O}$ for a given operator $O$ in the $S=1$ space. However, there is a natural choice for such a $\wtil{O}$ since $\msf{P}_{\mr{XY}}$ is a product of the norm non-preserving operator {$G$} and the natural embedding of a spin-1 into two fractional spins, $\msf{F}_{\mr{VBS}}$, as {introduced in Eq.~\eqref{eq:local map vbs}}. Namely, once one finds an operator $O'$ in the $S=1$ space that satisfies $OG=GO'$ (this is always possible since $G$ is bijective), $O'$ can be naturally fractionalized as a two-spin operator using Eq.~\eqref{eq:natural map}.

As a simple example, let us consider how $S^z_i$ can be fractionalized. Since $S_i^z$ commutes with $G$, we find
\begin{equation}
    S_i^zG\msf{F}_{\mr{VBS}}=GS_i^z\msf{F}_{\mr{VBS}}=G\msf{F}_{\mr{VBS}}\left(s_{2i-1}^z+s_{2i}^z\right),
\end{equation}
which implies Eq.~\eqref{eq:Sz fractional}.

The XY term $T_{i,i+1}\equiv S_i^xS^x_{i+1}+S_i^yS^y_{i+1}$ is more complex due to the non-trivial stretching transformation $G$ involved in teh embedding map. By commuting with $G$ we obtain
\begin{equation}\begin{split}
    T_{i,i+1}G&=GT'_{i,i+1}\\
    T'_{i,i+1}&\equiv\frac{1}{2}\ket*{0,0}\left(\bra*{+,-}+\bra*{-,+}\right)_{i,i+1}\\&+2\left(\ket*{+,-}+\ket*{-,+}\right)\bra*{0,0}_{i,i+1}\\
    &+\dyad*{+,0}{0,+}_{i,i+1}+\dyad*{0,+}{+,0}_{i,i+1}\\
    &+\dyad*{-,0}{0,-}_{i,i+1}+\dyad*{0,-}{-,0}_{i,i+1}.
    \end{split}
\end{equation}
The remaining task is to embed the operator $T'_{i,i+1}$ in the fractional spin space. One finds,
\begin{equation}
    \begin{split}
        T_{i,i+1}\msf{P}_{\mr{XY}}&=\msf{P}_{\mr{XY}}\wtil{T}_{i,i+1},
    \end{split}
\end{equation}
where 
\begin{widetext}
    \begin{equation}
        \begin{split}
            \wtil{T}_{i,i+1}&\equiv\frac{1}{4}\left(\ket*{\uparrow\downarrow\uparrow\downarrow}+\ket*{\uparrow\downarrow\downarrow\uparrow}+\ket*{\downarrow\uparrow\uparrow\downarrow}+\ket*{\downarrow\uparrow\downarrow\uparrow}\right)\left(\bra*{\uparrow\uparrow\downarrow\downarrow}+\bra*{\downarrow\downarrow\uparrow\uparrow}\right)_{2i-1;2i+2}\\
            &+\left(\ket*{\uparrow\uparrow\downarrow\downarrow}+\ket*{\downarrow\downarrow\uparrow\uparrow}\right)\left(\bra*{\uparrow\downarrow\uparrow\downarrow}+\bra*{\uparrow\downarrow\downarrow\uparrow}+\bra*{\downarrow\uparrow\uparrow\downarrow}+\bra*{\downarrow\uparrow\downarrow\uparrow}\right)_{2i-1;2i+2}\\
            &+\frac{1}{2}\left(\ket*{\uparrow\uparrow\uparrow\downarrow}+\ket*{\uparrow\uparrow\downarrow\uparrow}\right)\left(\bra*{\uparrow\downarrow\uparrow\uparrow}+\bra*{\downarrow\uparrow\uparrow\uparrow}\right)_{2i-1;2i+1}+\frac{1}{2}\left(\ket*{\uparrow\downarrow\uparrow\uparrow}+\ket*{\downarrow\uparrow\uparrow\uparrow}\right)\left(\bra*{\uparrow\uparrow\uparrow\downarrow}+\bra*{\uparrow\uparrow\downarrow\uparrow}\right)_{2i-1;2i+2}\\
            &+\frac{1}{2}\left(\ket*{\downarrow\downarrow\uparrow\downarrow}+\ket*{\downarrow\downarrow\downarrow\uparrow}\right)\left(\bra*{\uparrow\downarrow\downarrow\downarrow}+\bra*{\downarrow\uparrow\downarrow\downarrow}\right)_{2i-1;2i+1}+\frac{1}{2}\left(\ket*{\uparrow\downarrow\downarrow\downarrow}+\ket*{\downarrow\uparrow\downarrow\downarrow}\right)\left(\bra*{\downarrow\downarrow\uparrow\downarrow}+\bra*{\downarrow\downarrow\downarrow\uparrow}\right)_{2i-1;2i+2}.
        \end{split}
    \end{equation}
\end{widetext}
\subsection{Transforming operators to annihilators}
As mentioned in the main text, the bi-magnon scars are left invariant by the two-body projector $P^{2-\mr{body}}_{2i,2i+1}$. Thus it suffices to consider relevant terms in $\wtil{T}_{i,i+1}$, which are obtained by multiplying the operator from the right with the projector $P^{2-\mr{body}}_{2i,2i+1}$,
\begin{equation}\label{eq:rel T XY}
    \begin{split}
        \wtil{T}^{\mr{rel}}_{i,i+1}&\equiv\wtil{T}_{i,i+1}P_{2i,2i+1}^{2-\mr{body}}\\
        &=\left(\ket*{\uparrow\uparrow\downarrow\downarrow}+\ket*{\downarrow\downarrow\uparrow\uparrow}\right)\left(\bra*{\uparrow\downarrow\downarrow\uparrow}+\bra*{\downarrow\uparrow\uparrow\downarrow}\right)_{2i-1;2i+2}\\
        &+\frac{1}{2}\left(\ket*{\uparrow\uparrow\uparrow\downarrow}+\ket*{\uparrow\uparrow\downarrow\uparrow}\right)\bra*{\downarrow\uparrow\uparrow\uparrow}_{2i-1;2i+2}\\
        &+\frac{1}{2}\left(\ket*{\uparrow\downarrow\uparrow\uparrow}+\ket*{\downarrow\uparrow\uparrow\uparrow}\right)\bra*{\uparrow\uparrow\uparrow\downarrow}_{2i-1;2i+2}\\
        &+\frac{1}{2}\left(\ket*{\downarrow\downarrow\uparrow\downarrow}+\ket*{\downarrow\downarrow\downarrow\uparrow}\right)\bra*{\uparrow\downarrow\downarrow\downarrow}_{2i-1;2i+2}\\
        &+\frac{1}{2}\left(\ket*{\uparrow\downarrow\downarrow\downarrow}+\ket*{\downarrow\uparrow\downarrow\downarrow}\right)\bra*{\downarrow\downarrow\downarrow\uparrow}_{2i-1;2i+2}.
    \end{split}
\end{equation}

$\wtil{T}^{\mr{rel}}_{i,i+1}$ is not yet a sum of local annihilators. We therefore seek an operator equivalent to $\wtil{T}^{\mr{rel}}_{i,i+1}$ that does have that property. Since $(\ket*{\uparrow\downarrow}-\ket*{\downarrow\uparrow})_{2i-1,2i}$ is annihilated by $\msf{P}_{\mr{XY}}$, we can find, for example, that the term on the second line of Eq.~\eqref{eq:rel T XY} is equivalent to
\begin{equation}\begin{split}
    &\frac{1}{2}\left(\ket*{\uparrow\uparrow\uparrow\downarrow}+\ket*{\uparrow\uparrow\downarrow\uparrow}\right)\bra*{\downarrow\uparrow\uparrow\uparrow}_{2i-1;2i+2}\\
    &\sim_{\mr{XY}}\dyad*{\uparrow\uparrow\downarrow\uparrow}{\downarrow\uparrow\uparrow\uparrow}_{2i-1;2i+2}.
    \end{split}
\end{equation}
Using such properties, a suitable operator $\wtil{T}'_{i,i+1}$ equivalent to $\wtil{T}^{\mr{rel}}_{i,i+1}$ (i.e., $\wtil{T}'_{i,i+1}\sim_{\mr{XY}}\wtil{T}^{\mr{rel}}_{i,i+1}$) can be found in the form
\begin{equation}\begin{split}
    \wtil{T}'_{i,i+1}&=\left(\ket*{\uparrow\uparrow\downarrow\downarrow}+\ket*{\downarrow\downarrow\uparrow\uparrow}\right)\left(\bra*{\uparrow\downarrow\downarrow\uparrow}+\bra*{\downarrow\uparrow\uparrow\downarrow}\right)_{2i-1;2i+2}\\
    &\quad+\dyad*{\uparrow\uparrow\downarrow\uparrow}{\downarrow\uparrow\uparrow\uparrow}_{2i-1;2i+2}+\dyad*{\uparrow\downarrow\uparrow\uparrow}{\uparrow\uparrow\uparrow\downarrow}_{2i-1;2i+2}\\
    &\quad+\dyad*{\downarrow\downarrow\uparrow\downarrow}{\uparrow\downarrow\downarrow\downarrow}_{2i-1;2i+2}+\dyad*{\downarrow\uparrow\downarrow\downarrow}{\downarrow\downarrow\downarrow\uparrow}_{2i-1;2i+2}\\
    &=\dyad*{\uparrow\downarrow\downarrow}{\downarrow\downarrow\uparrow}_{2i;2i+2}+\dyad*{\downarrow\uparrow\uparrow}{\uparrow\uparrow\downarrow}_{2i;2i+2}\\
    &\quad+\dyad*{\uparrow\uparrow\downarrow}{\downarrow\uparrow\uparrow}_{2i-1;2i+1}+\dyad*{\downarrow\downarrow\uparrow}{\uparrow\downarrow\downarrow}_{2i-1;2i+1}.
    \end{split}
\end{equation}
In the last line a resolution of the identity was used to turn $\wtil{T}'_{i,i+1}$ into a three-body operator. We can further multiply $\wtil{T}'_{i,i+1}$ from the right by the two body operators $P_{2(i+k),2(i+k)+1}^{2-\mr{body}}$ (with {$k=-1,1$}) that overlap with the support of the individual terms 
so as to extract the part of the operator that acts non-trivially on the subspace to which the bi-magnon scars belong, 
\begin{equation}
    \begin{split}
        \wtil{T}''_{i,i+1}&{=\left(\dyad*{\uparrow\downarrow\downarrow}{\downarrow\downarrow\uparrow}+\dyad*{\downarrow\uparrow\uparrow}{\uparrow\uparrow\downarrow}\right)_{2i;2i+2}P^{2\mr{-body}}_{2i+2,2i+3}}\\
        &{+\left(\dyad*{\uparrow\uparrow\downarrow}{\downarrow\uparrow\uparrow}+\dyad*{\downarrow\downarrow\uparrow}{\uparrow\downarrow\downarrow}\right)_{2i-1;2i+1}P^{2\mr{-body}}_{2i-2,2i-1}}\\
        &{=}\dyad*{\uparrow\downarrow\downarrow\uparrow}{\downarrow\downarrow\uparrow\uparrow}_{2i;2i+3}+\dyad*{\downarrow\uparrow\uparrow\downarrow}{\uparrow\uparrow\downarrow\downarrow}_{2i;2i+3}\\
        &+\dyad*{\downarrow\uparrow\uparrow\downarrow}{\downarrow\downarrow\uparrow\uparrow}_{2i-2;2i+1}+\dyad*{\uparrow\downarrow\downarrow\uparrow}{\uparrow\uparrow\downarrow\downarrow}_{2i-2;2i+1}.
    \end{split}
\end{equation}
Summing these terms over all sites we finally obtain
\begin{equation}\begin{split}
    \sum_{i\in\Lambda}\wtil{T}''_{i,i+1}&=\sum_{i\in\Lambda}\ket*{\uparrow\downarrow\downarrow\uparrow}\left(\bra*{\uparrow\uparrow\downarrow\downarrow}+\bra*{\downarrow\downarrow\uparrow\uparrow}\right)_{2i;2i+3}\\
    &+\sum_{i\in\Lambda}\ket*{\downarrow\uparrow\uparrow\downarrow}\left(\bra*{\uparrow\uparrow\downarrow\downarrow}+\bra*{\downarrow\downarrow\uparrow\uparrow}\right)_{2i;2i+3},
    \end{split}
\end{equation}
each of which annihilates the bi-magnon scars $\ket*{\mc{B}_n^{\mr{XY}}}$, since $\wtil{T}''_{i,i+1}P_{2i;2i+3}^{4-\mr{body}}=0$ for any $i$. 

The sequence of transformations above shows explicitly that the suitably lifted XY term can be  rewritten as a sum of {\em local} terms, each of which annihilates the bi-magnon scar subspace. Since the bi-magnon states are eigenstates of the Zeeman term, this completes the proof that they are eigenstates of the full Hamiltonian.

\begin{widetext}
\section{Detailed proof of scar eigenstates in the AKLT model}\label{sec:aklt app}
Here we provide detailed proof that the scar states on the extended Hilbert space for the AKLT model are eigenstates of a suitably lifted Hamiltonian. As mentioned in the main text, it suffices to show that $\wtil{H}^>_{\mr{ann}}$ defined in Eq.~\eqref{eq:aklt decomp} is equivalent to a sum of local annihilators.
The projectors $\wtil{P}_{i,i+1}^{(2,1)}$ and $\wtil{P}^{(2,2)}_{i,i+1}$ which occur in $\wtil{H}_{\rm AKLT}$ have the following explicit form,
    \begin{equation}
    \begin{split}
        \wtil{P}_{i,i+1}^{(2,1)}&=\frac{1}{4}\left(\ket*{\downarrow\uparrow\uparrow\uparrow}+\ket*{\uparrow\downarrow\uparrow\uparrow}+\ket*{\uparrow\uparrow\downarrow\uparrow}+\ket*{\uparrow\uparrow\uparrow\downarrow}\right)\left(\bra*{\downarrow\uparrow\uparrow\uparrow}+\bra*{\uparrow\downarrow\uparrow\uparrow}+\bra*{\uparrow\uparrow\downarrow\uparrow}+\bra*{\uparrow\uparrow\uparrow\downarrow}\right)_{2i-1;2i+2}\\
        \wtil{P}_{i,i+1}^{(2,2)}&=\dyad*{\uparrow\uparrow\uparrow\uparrow}_{2i-1;2i+2}.
    \end{split}
\end{equation}
Like in the bi-magnon case, operators acting non-trivially on the scar states are obtained by multiplying them from the right by the projectors $\prod_{k=-1,0,1}P^{2-\mr{body}}_{2(i-k),2(i-k)+1}P^{6-\mr{body}}_{2i-2;2i+3}$ that stabilize the scar subspace. This reduces the above operators to 
\begin{equation}\begin{split}
    &T^{(2,1)}_{i,i+1}\equiv\wtil{P}_{i,i+1}^{(2,1)}\prod_{k=-1,0,1}P^{2-\mr{body}}_{2(i-k),2(i-k)+1}P_{2i-2;2i+3}^{6-\mr{body}}\\&=\frac{1}{4\sqrt{2}}\left(\ket*{\uparrow\downarrow\uparrow\uparrow\uparrow\uparrow}+\ket*{\uparrow\uparrow\downarrow\uparrow\uparrow\uparrow}+\ket*{\uparrow\uparrow\uparrow\downarrow\uparrow\uparrow}+\ket*{\uparrow\uparrow\uparrow\uparrow\downarrow\uparrow}\right)\left(\bra*{\Downarrow\Uparrow\Uparrow}-\bra*{\Uparrow\Uparrow\Downarrow}\right)_{2i-2;2i+3}\\    &T^{(2,2)}_{i,i+1}\equiv\wtil{P}_{i,i+1}^{(2,2)}\prod_{k=-1,0,1}P^{2-\mr{body}}_{2(i-k),2(i-k)+1}P_{2i-2;2i+3}^{6\mr{-body}}\\
    &=\dyad*{\uparrow\uparrow\uparrow\uparrow\uparrow\uparrow}{\Uparrow\Uparrow\Uparrow}_{2i-2;2i+3}-\frac{1}{\sqrt{2}}\dyad*{\downarrow\uparrow\uparrow\uparrow\uparrow\uparrow}{\Downarrow\Uparrow\Uparrow}_{2i-2;2i+3}+\frac{1}{\sqrt{2}}\dyad*{\uparrow\uparrow\uparrow\uparrow\uparrow\downarrow}{\Uparrow\Uparrow\Downarrow}_{2i-2;2i+3}.
    \end{split}
\end{equation}
Note that $T^{(2,1)}_{i,i+1}$ is equivalent to the following operator $T'^{(2,1)}_{i,i+1}$ (i.e., $T^{(2,1)}_{i,i+1}\sim_{\mr{VBS}}T'^{(2,1)}_{i,i+1}$),
\begin{equation}
    \begin{split}
        T'^{(2,1)}_{i,i+1}&=\frac{1}{2\sqrt{2}}\left(\ket*{\uparrow\downarrow\Uparrow\Uparrow}+\ket*{\Uparrow\Uparrow\downarrow\uparrow}\right)\left(\bra*{\Downarrow\Uparrow\Uparrow}-\bra*{\Uparrow\Uparrow\Downarrow}\right)_{2i-2;2i+3}.
    \end{split}
\end{equation}
Moreover, the Zeeman term $s^z_{2i}+s^z_{2i+1}$ reduces to following non-trivially acting term,
\begin{equation}
    \begin{split}
        T_{i,i+1}^{zz}&\coloneqq\left(s^z_{2i}+s^z_{2i+1}\right)\prod_{k=-1,0,1}P^{2-\mr{body}}_{2(i-k),2(i-k)+1}P^{6-\mr{body}}_{2i-2;2i+3}\\
        &=\dyad*{\Uparrow\Uparrow\Uparrow}_{2i-2;2i+3}+\dyad*{\Uparrow\Uparrow\Downarrow}_{2i-2;2i+3}+\dyad*{\Downarrow\Uparrow\Uparrow}_{2i-2;2i+3}.
    \end{split}
\end{equation}
\end{widetext}

Combining the three terms as they occur in the lifted Hamiltonian, one obtains the compact expression
\begin{equation}\begin{split}
    &T'^{(2,1)}_{i,i+1}+T^{(2,2)}_{i,i+1}-T_{i,i+1}^{zz}\\&=-\frac{1}{2\sqrt{2}}\left(\ket*{\uparrow\downarrow\Uparrow\Uparrow}-\ket*{\Uparrow\Uparrow\downarrow\uparrow}\right)\left(\bra*{\Uparrow\Uparrow\Downarrow}+\bra*{\Downarrow\Uparrow\Uparrow}\right)_{2i-2;2i+3}.
    \end{split}
\end{equation}
Each of these terms annihilates the scar subspace. Indeed their bra part is orthogonal to all possible 3-block configurations (Eq.~\eqref{eq:AKLT scar local}) in the scar states, mainly a consequence of the fact that pairs of neighboring block excitations are created with momentum $\pi$.
Upon summing the above operator over $i$, one obtains an operator equivalent to $\wtil{H}^>_{\mr{ann}}$. The above manipulations thus show how  $\wtil{H}^>_{\mr{ann}}$ can be transformed (up to equivalence) into a sum of local annihilators for the parent scar states.



\section{Similarities between scar states in constrained models and gauge theories}\label{app:gauge}
In the main text, we discuss the similarity between  scar states in constrained models and certain lattice gauge theories. The ground states of gauge theories in the deconfined phase are often expressed as simple low-entangled states projected onto the physical space defined by Gauss' law. Here we show how this works in the Ising gauge theory (IGT) and the quantum dimer model (QDM) and elucidate the similarity with our construction of scar states by Hilbert space enlargement and subsequent projection.
\begin{figure}
    \centering
    \includegraphics[width=.4\textwidth]{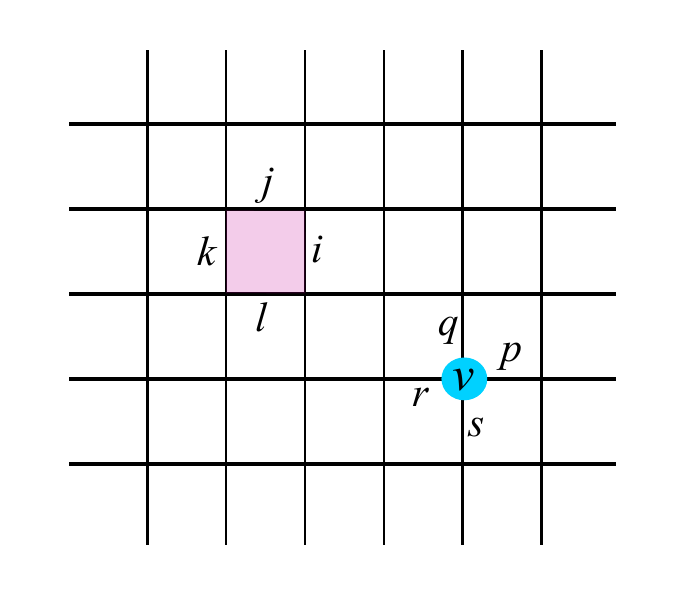}
    \caption{The square lattice with a plaquette $\square\in\mc{P}$ (the pink square) surrounded by the links $i,j,k$ and $l$, and with a vertex $v$ (the blue dot) surrounded by the links $p,q,r$, and $s$.}
    \label{fig:square lattice}
\end{figure}

\subsection{Deconfined phase in the Ising gauge theory}
To illustrate the {above} mentioned recipt {of constructing  ground states of gauge theories}, we first consider the IGT on the square lattice in its deconfined phase. The square lattice contains  links $\mc{E}=\mc{E}_{\mr{hor}}\cup\mc{E}_{\mr{ver}}$, where $\mc{E}_{\mr{hor}}(\mc{E}_{\mr{ver}})$ denotes the set of horizontal (vertical) links. A plaquette $\mc{P}$ is a set of four connected links forming a loop.
We assign Ising degrees of freedom $\sigma=\uparrow,\downarrow$ to every link of the square lattice. Links and plaquettes are denoted by $i$ and $\square$, respectively. We consider the following Hamiltonian $H_{\mr{IGT}}$,
\begin{equation}
    H_{\mr{IGT}}=-\Gamma\sum_{\square\in\mc{P}}\sigma^z_i\sigma^z_j\sigma^z_k\sigma^z_l,
\end{equation}
where the links $i,j,k$ and $l$ surround the plaquette $\square$ (see Fig.~\ref{fig:square lattice}). Furthermore, we impose the Gauss' law $G^{\mr{IGT}}_v$ for each vertex $v$,
\begin{equation}
    G^{\mr{IGT}}_v\coloneqq\sigma^x_p\sigma^x_q\sigma^x_r\sigma^x_s,
\end{equation}
where the links $p,q,r$ and $s$ emanate from the vertex $v$ (see Fig.~\ref{fig:square lattice}). The physical space is then defined as $\mc{V}^{\mr{IGT}}_{\mr{gauge}}\coloneqq\{\ket*{\psi}\large|\, G^{\mr{IGT}}_v\ket*{\psi}=\ket*{\psi}\,\,\forall v\}$. We define the corresponding projector onto $\mc{V}^{\mr{IGT}}_{\mr{gauge}}$ as $\msf{G}^{\mr{IGT}}_{\mr{gauge}}\coloneqq\prod_v(1+G^{\mr{IGT}}_v)/2$. The Gauss' law constraint $G^{\mr{IGT}}_v$ commutes with the Hamiltonian and thus, so does the projector $\msf{G}^{\mr{IGT}}_{\mr{gauge}}$. In the unconstrained Hilbert space $H_{\mr{IGT}}$ has many degenerate ground states. It is easy to construct one of them, for example the product state
\begin{equation}\label{eq:gs IGT}
    \ket*{\psi^{\mr{IGT}}_{\mr{gs}}}\coloneqq\bigotimes_{i\in\mc{E}}\ket*{\uparrow}_i.
\end{equation}
If one fixes the axial gauge (the subspace constrained by the requirement $\sigma^z_i=1$ for $\forall i\in\mc{E}_{\mr{hor}}$) and moreover imposes $\sigma^z_j=1$ along one vertical line, this state is the unique ground state.
By projecting it onto the physical subspace $\mc{V}^{\mr{IGT}}_{\mr{gauge}}$, one then obtains the gauge invariant ground state, i.e., $\msf{G}^{\mr{IGT}}_{\mr{gauge}}\ket*{\psi^{\mr{IGT}}_{\mr{gs}}}$. 
The physical ground state wavefunction can be interpreted as an equal amplitude superposition of all possible configurations in which up-spins form closed loops. 

Note that the above construction of the IGT ground state is very similar to the construction of the scar states we discussed. Indeed, the relation $H_{\mr{IGT}}\msf{G}^{\mr{IGT}}_{\mr{gauge}}\ket*{\psi_{\mr{gs}}^{\mr{IGT}}}=\msf{G}^{\mr{IGT}}_{\mr{gauge}}H_{\mr{IGT}}\ket*{\psi^{\mr{IGT}}_{\mr{gs}}}$, can be interpreted as stating that the Hamiltonian lifted  to the extended, unconstrained space is identical to the original Hamiltonian in the physical subspace - a special case of the general situation we considered for scars. Moreover, the simple ground state found in the extended space is then projected down to a more complex wavefunction, similarly as rather simple scar wavefunctions in extended space get projected into more complex scars in the constrained space.  

On a surface of non-trivial genus $g$, $H_{\mr{IGT}}$  hosts $2^{2g}$ nearly degenerate ground states.
If, as in the model written down, 
there are exactly conserved string operators along non-contractable loops, one can define the associated projectors $P_{\mr{top}}$ onto a given topological sector, and those commute with the Hamiltonian.
The ground state in such a sector is then obtained upon further projecting  the state $\msf{G}^{\mr{IGT}}_{\mr{gauge}}\ket*{\psi^{\rm IGT}_{\mr{gs}}}$ onto the chosen sector. Indeed, it is easy to see that  $P_{\mr{top}} \msf{G}^{\mr{IGT}}_{\mr{gauge}}\ket*{\psi^{\rm IGT}_{\mr{gs}}}$ is a ground state, too. 

The fact that a given parent state maps to ground states in different topological sectors bears a strong similarity to how we generated the pyramid states $\ket*{\mc{S}^{\mr{Pyr}}_{n,m}}$. There the physical scar states were obtained by projecting the parent states onto the various sectors with a given conserved domain wall number.

\subsection{The Rokhsar-Kivelson point in the quantum dimer model}
The quantum dimer model (QDM) on the square lattice hosts a spin-liquid ground state at the so-called Rokhsar-Kivelson (RK) point. Following the convention of the IGT, we again assign Ising degrees of freedom $\sigma=\uparrow,\downarrow$ to every bond of the lattice, which now indicates the presence or absence of a dimer. The physical space consists in complete dimer coverings, where each vertex belongs to one and only one dimer. This hardcore condition can be represented by Gauss' law, $G^{\mr{QDM}}_v = 1$ for each vertex,
 $\mc{V}^{\mr{QDM}}_{\mr{gauge}}\coloneqq\{\ket*{\psi}\large|\,G^{\mr{QDM}}_v\ket*{\psi}=\ket*{\psi}\,\forall v\}$, where
\begin{equation}
    G^{\mr{QDM}}_v\coloneqq n_p+n_q+n_r+n_s,
\end{equation}
with the links $p,q,r,s$ attached to the vertex $v$.
 The global projector on the physical space is denoted by $\msf{G}^{\mr{QDM}}_{\mr{gauge}}=\prod_v  G^{\mr{QDM}}_v$.

The dynamics of the QDM is governed by the Hamiltonian, 
\begin{equation}\label{eq:QDM H}\begin{split}
    H_{\mr{QDM}}&=-\sum_{p\in\mc{P}}\left(\sigma^+_i\sigma^-_j\sigma^+_k\sigma^-_l+h.c.\right)\\
    &+\sum_{p\in\mc{P}}\left(z^2n_iP_jn_kP_l+\frac{1}{z^2}P_in_jP_kn_l\right)\\
    ={z^{-2}}\sum_{p\in\mc{P}}&{\left(z^2\ket*{\uparrow\downarrow\uparrow\downarrow}-\ket*{\downarrow\uparrow\downarrow\uparrow}\right)\left(z^2\bra*{\uparrow\downarrow\uparrow\downarrow}-\bra*{\downarrow\uparrow\downarrow\uparrow}\right)_{ijkl},}
    \end{split}
\end{equation}
where we recall that $P_i \equiv\dyad*{\downarrow}_i=1-n_i$. The links $i,j,k$ and $l$ surround the plaquette $p$, and $z$ is a real parameter. Note that dynamics preserves the dimer constraint. 

A simple ground state of the Hamiltonian on the unconstrained space is given by the product state 
\begin{equation}
    \ket*{\psi^{\mr{QDM}}_{\mr{gs}}}\coloneqq\bigotimes_{i\in\mc{E}_{\mr{hor}}}\left(\ket*{\uparrow}+\ket*{\downarrow}\right)_i\bigotimes_{j\in\mc{E}_{\mr{ver}}}\left(z\ket*{\uparrow}+\ket*{\downarrow}\right)_j.
\end{equation}
Indeed, this state is annihilated by each local projector {in the last line of Eq.~\eqref{eq:QDM H}} in $H_{\mr{QDM}}$ (one can check that the product state on a plaquette contains the two relevant antiferromagnetic patterns with weights 1 and $z^2$, respectively). This product state is one of the ground states in the subspace satisfying the gauge fixing constraint $\sigma_i^x=+1$ for all $i\in\mc{E}_{\mr{hor}}$.
A gauge invariant ground state is then again obtained by projection onto the physical subspace, $\msf{G}^{\mr{QDM}}_{\mr{gauge}}
\ket*{\psi^{\mr{QDM}}_{\mr{gs}}}$. {It is a weighted superposition of all possible dimer configurations.}
On a surface with non-trivial genus the ground state manifold is again topologically degenerate, 
and the ground state in each topological sector is obtained by appropriate projection.

\subsection{Similarity and difference from QMBS}
The ground states of  the IGT and the QDM share a similar structure with the scar states {in the constrained models we} discussed in the main text. Namely, in suitably enlarged (unconstrained) Hilbert spaces one can find simple expressions for these states, while the physical states are obtained upon projection onto the constrained Hilbert space. 

In gauge theories, 
finding the ground state in the enlarged, unconstrained Hilbert space becomes rather simple if a proper gauge is fixed.

We may re-interpret our scar construction for the AKLT model from this viewpoint. Instead of the original $S=1$ AKLT model, we consider the natural corresponding Hamiltonian $\wtil{H}_{\mr{AKLT}}$ (Eq.~\eqref{eq:aklt frac}) defined in the enlarged $S=1/2$ space. We define the projector $P^{S=1}_{2i-1,2i}$, which projects two spin-1/2s at the sites $2i-1$ and $2i$ onto the triplet sector. This projector is essentially equivalent to the map $F^{\mr{VBS}}_i$ defined in Eq.~\eqref{eq:local map vbs}. We can interpret the Hamiltonian $\wtil{H}_{\mr{AKLT}}$ as being equipped with an analog of the Gauss' law $P^{S=1}_{2i-1,2i}\ket*{\psi}=+\ket*{\psi}$ for $\forall i$. As $\wtil{H}_{\mr{AKLT}}$ commutes with $P^{S=1}_{2i-1,2i}$, one can find ``{physical}'' eigenstates, that is, states that live entirely in the spin 1 sector. The ground state in the enlarged space is expressed as a simple product state of singlets. This is the analog of the ground state in the axial gauge for the IGT. The physical ground state is then simply obtained by applying the projectors $P^{S=1}_{2i-1,2i}$.

We may add terms to the Hamiltonian that are non-zero only on  Gauss' law-violating configurations. While this does not alter its action  on the physical space, 
such gauge fixing terms lift the degeneracy of the ground states on the unconstrained space and thus favor a certain ``gauge''. For example, for the IGT the following additions impose the axial gauge on the ground state in the extended Hilbert space
\begin{equation}
    \begin{split}
        &H_{\mr{IGT}}\msf{G}^{\mr{IGT}}_{\mr{gauge}}=\msf{G}^{\mr{IGT}}_{\mr{gauge}}H_{\mr{IGT}}\\
        &=\msf{G}^{\mr{IGT}}_{\mr{gauge}}\left(H_{\mr{IGT}}+\frac{1}{2\xi}\sum_{i\in\mc{E}_{\mr{hor}}}\left(1-G_{v_i}^{\mr{IGT}}\right)\left(1-\sigma^z_i\right)\right)\\
        &\equiv\msf{G}^{\mr{IGT}}_{\mr{gauge}}\wtil{H}_{\mr{IGT}}.
    \end{split}
\end{equation}
Here $\xi$ is a positive constant and $v_i$ is any vertex which does not belong to the link $i\in\mc{E}_{\mr{hor}}$.

In the AKLT model the analog of this gauge fixing procedure corresponds to modifying $\wtil{H}^>_{\mr{ann}}$ in Eq.~\eqref{eq:aklt decomp} by adding terms that are annihilated by $P^{S=1}_{2i-1,2i}$. While such terms  have no effect on the gauge-invariant subspace their addition may bring the Hamiltonian into projector embedding form, such that scar states assume a simple structure and can be easily constructed.

We can also find an analog of the topological degeneracy of gauge theory ground states in the realm of QMBS.  
Topological sectors are dynamically disconnected subspaces. In cases where they are distinguished by t'Hooft operators that commute with the Hamiltonian one can project onto the various topological sectors, but the corresponding projectors are non-local. Simple ground states (in a suitable fixed gauge) are usually superpositions spanning all topological sectors, while the ground state within each sector is obtained upon further projection. 
This structure finds an analog in the family of towers of scar states in the domain wall model.
There the domain wall number discussed in Sec.~\ref{sec:dw model} takes the role of a global constraint (conservation law), which now takes the role of a string operator defining a topological sector. Accordingly, each domain wall number is associated with a separate tower of states. However, unlike ground states in different topological sectors, the associated towers of states are not energetically degenerate. This is because the domain walls come with an energetic cost, which shifts the scar spectra of different domain wall sectors by a constant with respect to each other. 

\end{document}